\documentclass[aps,prd,notitlepage,showpacs,nofootinbib,superscriptaddress]{revtex4-2}
\pdfoutput=1

\usepackage[usenames,dvipsnames]{xcolor}  
\usepackage{colortbl} 
\usepackage{graphicx} 
\definecolor{lightblue}{rgb}{0.93, 0.95, 1.0} 
\definecolor{lightgray}{gray}{0.9} 
\usepackage[compat=1.1.0]{tikz-feynman}
\usepackage{bm}
\usepackage{bbold}
\usepackage{latexsym}
\usepackage{amsthm}
\usepackage{amsmath}
\usepackage{amssymb}
\usepackage{slashed}
\usepackage{xfrac}
\usepackage{wrapfig}
\usepackage[justification=raggedright,singlelinecheck=false]{caption}

\usepackage{cancel}
\usepackage{multirow}
\usepackage{pst-3d}                  
\usepackage{pst-grad}                
\usepackage{pst-node}                
\usepackage{pst-slpe}                
\usepackage{pst-plot}                %
\usepackage{pst-coil}                %
\usepackage{pst-math}                %
\usepackage{pstricks-add}            %
\usepackage{rotating}
\usepackage{geometry}
\usepackage{hhline}
\usepackage{float}
\usepackage{fancybox}
\usepackage[utf8]{inputenc}
\usepackage{orcidlink}
\usepackage{array}
\newcolumntype{C}[1]{>{\centering\arraybackslash}p{#1}}

\usepackage{hyperref}
\hypersetup{
    colorlinks=true,
    citecolor=darkred,
    urlcolor=darkred,
    pdfborder={0 0 0}}
\usepackage[normalem]{ulem}

\usepackage{color,soul}
\setul{0.5ex}{0.3ex}
\setulcolor{blue}

\usepackage{braket}
\definecolor{linkcolor}{rgb}{0,0,0.5}
\allowdisplaybreaks
\allowdisplaybreaks[1]
\allowdisplaybreaks[2]

\definecolor{greenLinks}{rgb}{0, 0.6, 0}
\definecolor{blueLinks}{rgb}{0, 0, 0.6}
\definecolor{redLinks}{rgb}{0.6, 0, 0}
\definecolor{tempText}{rgb}{0.55, 0.10,0.67}
\definecolor{eprintLinks}{rgb}{0.4, 0.4, 0.4}
\definecolor{journalLinks}{rgb}{0.6, 0, 0}
\newcommand {\ignore}[1]{}

\usepackage{soul}
\definecolor{mightnightblue}{RGB}{25,25,112}
\definecolor{brown}{rgb}{0.59, 0.29, 0.0}
\definecolor{darkred}{rgb}{0.6,0,0}

\def\lsim{\mathrel{\rlap{\lower4pt\hbox{\hskip1pt$\sim$}}
    \raise1pt\hbox{$<$}}}
\def\gsim{\mathrel{\rlap{\lower4pt\hbox{\hskip1pt$\sim$}}
    \raise1pt\hbox{$>$}}}

\newcommand{\AddrIFIC}{%
  Instituto de F\'{i}sica Corpuscular, CSIC-Universitat de Val\`{e}ncia, 46980 Paterna, Spain\vspace{0.1cm}}
 
\newcommand{\AddrMPI}{Max-Planck-Institut f\"ur Kernphysik, Saupfercheckweg 1, 69117 Heidelberg, Germany\vspace{0.1cm}}

\newcommand{\AddrKIT}{Institut f\"ur Astroteilchenphysik, Karlsruher Institut f\"ur Technologie (KIT), Hermann-von-Helmholtz-Platz 1, 76344 Eggenstein-Leopoldshafen, Germany}

\begin{document}

\title{\boldmath \color{BrickRed} Testing lepton non-unitarity with the next generation of Germanium-based CE$\nu$NS reactor experiments}

\author{Salvador Centelles Chuli\'{a}\orcidlink{0000-0001-6579-1067}}\email{salcen@ific.uv.es}
\affiliation{\AddrIFIC}
\affiliation{\AddrMPI}
\author{Manfred Lindner\orcidlink{0000-0002-3704-6016}}\email{lindner@mpi-hd.mpg.de}
\affiliation{\AddrMPI}
\author{Thomas Rink\orcidlink{0000-0002-9293-1106}}\email{thomas.rink@kit.edu}
\affiliation{\AddrKIT}
\affiliation{\AddrMPI}


\begin{abstract}

\noindent 
Coherent elastic neutrino-nucleus scattering (CE$\nu$NS) has been experimentally confirmed using neutrinos from pion decay at rest, solar neutrinos and reactor antineutrinos. Future CE$\nu$NS experiments will foreseeable lead to precision measurements which will be a powerful tool to search for new physics beyond the Standard Model. In this work, we investigate possible deviations from unitarity in the $3\times3$ leptonic mixing matrix that controls the propagation of active neutrinos. Such deviations may originate from the mixing with additional gauge singlet fermions and depending on their mass scale and mixing, the resulting phenomenology can differ substantially. We explore two well-motivated regimes: the \textit{seesaw limit}, where the new fermions are heavy and kinematically inaccessible, leading to effective deviations from unitarity in the active sector; and the \textit{light sterile limit}, where they are light enough to be produced and participate in neutrino propagation and scattering processes. We show how these scenarios modify both CE$\nu$NS and elastic neutrino-electron scattering (E$\nu$eS), and we present the corresponding sensitivity projections for a future CE$\nu$NS reactor experiment obtained by upscaling the CONUS+ experiment, which reported the first observation of reactor CE$\nu$NS. We identify the leading experimental systematics relevant for such an upscaling and demonstrate the resulting capability to probe TeV-scale new physics. Our results highlight the strong potential of CE$\nu$NS to test the structure of the lepton sector and to search for physics beyond the Standard Model.

\end{abstract}


\maketitle


\section{Introduction}

\label{sec:intro}

\noindent The discovery of neutrino oscillations~\cite{Kajita:2016cak,McDonald:2016ixn} implies physics Beyond Standard Model (BSM), with massive and non-degenerate neutrinos being the simplest explanation. This has led to many projects and ideas mostly driven by experiments involving charged current (CC) processes aiming at improved determinations of neutrino masses and mixings. New physics may, however, manifest itself also in modified CC interactions and especially also in neutral current (NC) processes. It may also show up as unitarity violation as a consequence of new, gauge singlet fermions and the associated phenomena have been extensively explored~\cite{Escrihuela:2015wra,Escrihuela:2016ube,Fong:2016yyh,Ge:2016xya,Miranda:2016ptb,Miranda:2016wdr,Fong:2017gke,Miranda:2019ynh,Miranda:2020syh,Martinez-Soler:2021sir,Rahaman:2021cgc,Soumya:2021dmy,Kaur:2021rau,Wang:2021rsi,Chatterjee:2021xyu,Gariazzo:2022evs,Aloni:2022ebm,Sahoo:2023mpj,Celestino-Ramirez:2023zox,Miranda:2021kre,Schwetz:2021thj,Schwetz:2021cuj,Tang:2021lyn,Arrington:2022pon,Capozzi:2023ltl,Soleti:2023hlr}.
Another potentially powerful tool are coherently enhanced NC processes, in particular \textit{coherent elastic neutrino-nucleus scattering} - (CE$\nu$NS), pioneered by Freedman in the 70s~\cite{Freedman:1973yd}. CE$\nu$NS has been recently observed with pion decay-at-rest neutrinos~\cite{COHERENT:2017ipa,zenodo,COHERENT:2020iec,COHERENT:2024axu},
solar neutrinos~\cite{XENON:2024ijk, PandaX:2024muv, LZ:2025igz} and reactor antineutrinos~\cite{Ackermann:2025obx} which provides an interesting and complementary way to study neutrinos~\cite{Barranco:2005yy}. In this paper we analyze the potential of precision CE$\nu$NS measurements achievable with an upscaling of the successful CONUS+ technology.

Gauge singlet fermions are in general well motivated and three right-handed neutrinos are in fact one of the best ways to explain neutrino masses. This allows Dirac and Majorana mass terms leading to the famous see-saw mechanism. Diagonalization of the total mass matrix leads then usually to three light Majorana neutrinos, where their mixing matrix is unitary if the right-handed Majorana states are assumed to be ultra heavy. The right-handed states lead, however, to unitarity violations if they are not so heavy. Unitarity violations also show up if more gauge singlet fermions exist. The structure of both the CC and NC changes in all these cases in a non-trivial way~\cite{Schechter:1980gr,Schechter:1981cv}.
Indeed, the mixing matrix underlying the leptonic CC weak interaction that describes oscillations~\cite{deSalas:2020pgw} deviates from unitarity, while the NC interaction of mass eigenstate neutrinos deviates from the unit matrix, with these two features inter-connected.
Although the effect of non-unitary neutrino mixing was first discussed in the context of astrophysical neutrino propagation~\cite{Valle:1987gv,Nunokawa:1996tg,Grasso:1998tt}, it can be phenomenologically relevant in Earth-bound experiments. 

This happens in the context of genuine low-scale seesaw schemes, such as the inverse~\cite{Mohapatra:1986bd,GonzalezGarcia:1988rw} or the linear seesaw mechanism~\cite{Akhmedov:1995ip,Akhmedov:1995vm,Malinsky:2005bi}, leading to potentially sizable deviations from the conventional leptonic weak currents with unitary mixing.
These corrections are expressed as power series in the parameter $\varepsilon = \mathcal O(Y v / M)$, where $M$ is the mediator mass scale and $v$ is the SM vacuum expectation value (vev).
Although small, we stress that $\varepsilon$ can be non-negligible within low-scale realizations of the seesaw. We call this scenario \textit{the seesaw limit}, where the new states are heavy enough to decouple from low energy experiments while the mixing with active neutrinos remains sizable and not suppressed by neutrino mass. In what follows, we remain agnostic about the neutrino mass generation and consider also the \textit{light sterile limit}, where the gauge singlets are light enough to be produced and potentially participate in the neutrino propagation and scattering.

Here we explore the potential of CE$\nu$NS and elastic neutrino-electron scattering (E$\nu$eS) to probe non-minimal charged-current and neutral-current weak interactions within the context of CE$\nu$NS experiments using Germanium detector technology close to a nuclear power reactor. Motivated by the recent detection of CE$\nu$NS by the CONUS+ experiment~\cite{Ackermann:2025obx} and the established role Germanium technology occupies in this field \cite{CONUS:2020skt,Bonet:2021wjw,CONUS:2021dwh,nGeN:2022uje,CONUSCollaboration:2024kvo,TEXONO:2024vfk,Yang:2024exl,nGeN:2025hsd,CONUS:2024lnu} we want to access the future potential of such experiments. At the same time, further experiments using other detection technologies are catching up, among them: CONNIE~\cite{CONNIE:2021ggh}, NEON~\cite{NEON:2022hbk}, NUCLEUS~\cite{NUCLEUS:2019igx}, RED-100~\cite{RED-100:2024izi}, RELICS~\cite{RELICS:2024opj} and Ricochet~\cite{Ricochet:2021rjo}. 
More insightful data can be also be expected from the beam side~\cite{Baxter:2019mcx,Simon:2024cat} and dark matter direct detection experiments~\cite{XENON:2024ijk,PandaX:2024muv}, where solar neutrinos have already been observed.
First data provided by these experiments already allowed for various CE$\nu$NS investigations within and beyond the SM such as determinations of the Weinberg angle at these low energies, searches for non-standard neutrino interactions (NSI), light (mediator) particles or electromagnetic properties of the neutrino~\cite{Coloma:2017ncl,Papoulias:2017qdn,Liao:2017uzy,Abdullah:2018ykz,AristizabalSierra:2018eqm,Khan:2019cvi,Miranda:2020syh,Cadeddu:2020lky,Miranda:2020tif,Abdullah:2022zue,Huang:2025znh}. Our study complements recent theoretical works that investigate both SM and BSM scenarios using Germanium reactor data \cite{AtzoriCorona:2025xgj,AtzoriCorona:2025ygn,DeRomeri:2025csu,Chattaraj:2025fvx,Alpizar-Venegas:2025wor}, and aims to shed light on the scale of neutrino mass generation.

The paper is organized as follows: In Section \ref{sec:theory} we introduce the theoretical framework we apply throughout this work and show how CE$\nu$NS and E$\nu$eS are altered by non-unitarity effects. In Section \ref{sec:experiment} we analyze the potential of a future experiment which we choose to be an upscaling of the successful CONUS+ technology to larger detector masses. Our findings are presented in Section \ref{sec:results} for the cases of light and heavy new physics and for selected experimental specifications. We summarize our findings and conclude with Section \ref{sec:conclusion}.


\section{Theory preliminaries} \label{sec:theory}


\noindent We assume that the $3$ active neutrinos mix with $m$ gauge singlet fermion states. In this case, the most general charged current weak interaction of massive neutrinos is described by a rectangular matrix $K$~\cite{Schechter:1980gr}, connecting $3$ charged lepton mass states with $3+m$ massive neutral states:

\begin{eqnarray}
-\mathcal{L}_{\mathrm{CC}} = \frac{g}{\sqrt{2}} W^+_\mu  \bar{\ell}_L \, \gamma^\mu \,  K \, \nu    + h.c.\, ,
\end{eqnarray}
with $\bar{\ell}_L = (\bar{e}_L, \bar{\mu}_L, \bar{\tau}_L)$ and $\nu = (\nu_1, \nu_2, ... \nu_{3+m})^T$ the mass eigenstates of the charged leptons and neutrinos, respectively. $K$ is a rectangular $3 \times (3+m)$ matrix, where in the basis where the charged lepton mass matrix is diagonal, the upper blocks are simply the first 3 rows of the neutrino mixing matrix~\cite{Schechter:1981cv}. We can also define the relevant sub-block as
\begin{equation}
    K = \left(\begin{matrix} N & S \end{matrix}\right) \label{eq:KNS}\, ,
\end{equation}
with $N$ a $(3\times 3)$ matrix, while $S$ is $ (3\times m)$. Further $K K^\dagger = I_{3\times 3}$ and thus $S S^\dagger = I - N N^\dagger$, but note that in general $P = K^\dagger K \neq I$. The matrix $P$, which is square $(3+m)\times (3+m)$, non-diagonal, non-unitary and projective ($P^2= P$)  parametrizes the neutral current interaction given by

\begin{eqnarray}
    -\mathcal{L}_{NC} = \frac{g}{2 \cos \theta_W} \bar{\nu} \, \gamma^\mu P_L K^\dagger K  \nu = \frac{g}{2 \cos \theta_W} \bar{\nu} \, \gamma^\mu P_L \, P \, \nu\, .
\end{eqnarray}

\noindent This formalism applies to both heavy and light singlet states. In what follows, we will now distinguish between two relevant limits, depending on the mass of the new neutral leptons.

\subsection{Heavy neutral leptons: the seesaw limit}

\noindent We first assume $m_4, m_5, ... \sim M \gg \Lambda_{\mathrm{EW}}$, which we call the seesaw limit. 
While our phenomenological analysis remains model-agnostic, it is useful to recall what is theoretically expected. 
In seesaw scenarios, the hierarchy between the heavy singlet states and the light active ones allows for an expansion in a small parameter $\varepsilon$, which quantifies the departure from unitarity in the active sector. 
In canonical high-scale seesaws, such as the type-I mechanism, this parameter scales as $\varepsilon \sim \mathcal{O}(\sqrt{m_\nu/M}) \ll 1$ and therefore leads to unobservable effects. 
By contrast, in genuine low-scale realizations, paradigmatically the inverse or linear seesaws, the expansion parameter is $\varepsilon \sim \mathcal{O}(m_D/M)$, with $m_D = Yv$ the Dirac mass generated by the Yukawa interaction of coupling $Y$ and $v$ the Higgs vacuum expectation value. 
Assuming $Y = \mathcal{O}(1)$, any experimental bound on $\varepsilon$ can thus be interpreted as a direct constraint on the heavy-mediator mass scale $M$. 
Current global fits to oscillation data~\cite{Forero:2021azc} typically require $\varepsilon^2 \lesssim \mathcal{O}(10^{-2})$, corresponding to $M \gtrsim 1.7\,\mathrm{TeV}$ for $Y \sim 1$.

If the energy of a given process is well below the mass of the heavy mediators, they cannot be produced and therefore do not participate in oscillation experiments.
Then, effectively, only the first $3\times3$ blocks of $K$ and $P$ will play a role in the weak interactions, i.e.\ $N$ in the charged current and $N^\dagger N$ in the neutral current. 
We can relate the order of these blocks with the seesaw expansion parameter $\varepsilon \sim \mathcal{O}(m_D/M)$ as
\begin{equation}
N N^\dagger \sim 1-\mathcal{O}\left(\varepsilon^2\right), 
\hspace{1cm}   
S S^\dagger \sim \mathcal{O}\left(\varepsilon^2\right)\, .
\end{equation}

The resulting non-unitary matrix $N$ can be parametrized following~\cite{Escrihuela:2015wra} as~\footnote{An alternative description and its relationship with Eq.\ \ref{eq:Nparam} is discussed in Ref.~\cite{Blennow:2016jkn}.}
\begin{eqnarray}
\label{eq:Nparam}
  N &=& \left(\begin{matrix}
      \alpha_{11} &     0       & 0 \\
      \alpha_{21} & \alpha_{22} & 0 \\
      \alpha_{31} & \alpha_{32} & \alpha_{33}
  \end{matrix}\right) \cdot U \, .
\end{eqnarray}
 Besides the $3\times 3$ unitary matrix $U$ used to describe neutrino mixing in the conventional unitary case, one has the triangular prefactor characterized by 3 real diagonal $\alpha_{ii}$ ($i=1,2,3$), and 3 non-diagonal $\alpha_{ij}$ ($i\neq j$) which are complex.
 Note that the $\alpha_{ij}$ parameters have a direct interpretation in terms of the mixing angles between active neutrinos and the heavy singlet states. As an illustrative example, in the $3+1$ scheme with one heavy neutral lepton we find $\alpha_{ii} = \cos \theta_{i4}$ for the diagonal entries, while the off-diagonal ones are given by $\alpha_{ij} = \sin \theta_{i4} \sin \theta_{j4} e^{i (\phi_{i4}-\phi_{j4})}$, where $\theta_{i4}$ and $\phi_{i4}$ are the mixing angle and phases between the active and sterile sectors. Since the mixing angles are expected to be small, the diagonal terms are close to $1$ and real while the off-diagonal entries are small and complex. The general expressions for an arbitrary number of heavy neutral leptons can be found in the appendix of \cite{Escrihuela:2015wra}.
This is a convenient and complete description of non-unitarity in the lepton sector.
By construction, $N$ and $S$ must satisfy the relation $N N^\dagger = 1 - S S^\dagger$, hence $N N^\dagger \sim 1 - \mathcal{O}\left(\varepsilon^2\right)$. Explicitly,
\begin{equation}
    N N^\dagger  = \begin{pmatrix}
        \alpha_{11}^2           & \alpha_{11} \alpha_{21}^*         & \alpha_{11} \alpha_{31}^* \\
        \alpha_{11} \alpha_{21} & \alpha_{22}^2 + |\alpha_{21}|^2   & \alpha_{22} \alpha_{32}^* + \alpha_{21} \alpha_{31}^* \\
        \alpha_{11} \alpha_{31} & \alpha_{22} \alpha_{32} + \alpha_{21}^* \alpha_{31}          & \alpha_{33}^2 + |\alpha_{31}|^2 + |\alpha_{32}|^2
    \end{pmatrix}~,
\end{equation}
from which one can read off the strength of the $\alpha_{ij}$ in terms of the small seesaw expansion parameter $\varepsilon$:
\begin{eqnarray}
   \label{eq:epsorder1} \alpha_{ii}^2 &\sim& 1-\mathcal{O}\left(\varepsilon^2\right)\, , \\
    \label{eq:epsorder2} |\alpha_{ij}|^2 &\sim& \mathcal{O}\left(\varepsilon^4\right) , \hspace{0.5cm} i\neq j~.
\end{eqnarray}
One sees that the strength of the off-diagonal $\alpha$ parameters is suppressed relative to the deviations of the flavour-diagonal ones from their SM values. 
In zero-distance experiments, where neutrinos cannot oscillate from the source to the detector, the $0^\text{th}$ order in the seesaw expansion corresponds to the unitary limit, the $1^\text{st}$ order gives only diagonal flavour-conserving effects, while
the genuine flavour-violating effects of non-unitary only appear at $2^\text{nd}$ order. Notice also that this behavior is consistent with the validity of the well-known triangle inequality $|\alpha_{ij}| \leq \sqrt{(1-\alpha^2_{ii})(1-\alpha^2_{jj})}$~\cite{Escrihuela:2015wra}.

From Eq.\ \ref{eq:epsorder1} a bound on $1-\alpha_{ii}^2$ can be interpreted, up to $\mathcal{O}(1)$ factors, as a bound on $\varepsilon^2$. Using $\varepsilon \sim m_D/M$ and $m_D = Yv$, one obtains the illustrative relation
\begin{eqnarray}\label{eq:limit_to_mass}
    M \gtrsim v \frac{1}{\sqrt{(1-\alpha_{ii}^2)^{\text{limit}}}} \, .
    \label{eq:Mlimit}
\end{eqnarray}
Here $(1-\alpha_{ii}^2)^{\text{limit}}$ denotes the experimental upper bound on the non-unitarity parameter $(1-\alpha_{ii}^2)$. Eq.\ \ref{eq:Mlimit} should not be interpreted as a strict experimental constraint on $M$, but rather as an intuitive translation between the $\alpha$ parameters and the characteristic mass scale of low-scale seesaw mediators, assuming $\mathcal{O}(1)$ Yukawa couplings.

Additionally, unitarity violation leads to a redefinition of the Fermi constant, which is extracted from the muon lifetime assuming the SM to be valid.
In the presence of non-unitarity the measured quantity would be an effective muon decay coupling $G_\mu$. Since the W boson vertices are modified by the non-unitarity parameters one finds
\begin{align}
G_\mu &= 1.1663787(6) \times 10^{-5} \text{\,GeV}^{-2} & \text{(effective $\mu^-$ decay constant \cite{Workman:2022ynf})} \\
G^2_\mu &= G^2_F (N N^\dagger)_{ee} (N N^\dagger)_{\mu \mu}
\end{align}
and therefore
\begin{equation}
\label{eq:GF}
    1 \leq \frac{G_F^2}{G_\mu^2} = \frac{1}{(N N^\dagger)_{ee} (N N^\dagger)_{\mu \mu}} \approx 3 - \alpha_{11}^2 - \alpha_{22}^2 \sim 1+ \mathcal{O}\left(\varepsilon^2\right)\, .
\end{equation}

Consequently, any process proportional to $G_F^2$ will receive an ``enhancement'' due the deviation from unitarity. 
This is counterintuitive because naively one expects less events than in the SM if the mixing is non-unitary, due to kinematically inaccessible heavy states.  
The reduction of the event number due to non-unitarity, and the ``increase'' due to the redefinition of $G_F$ compete with each other, 
so that in some cases one can achieve $\mathcal{N}_\text{SM}/\mathcal{N}_\text{NU} = 1$ even in the presence of non-unitarity.

In the context of future CONUS-like experiments, we are interested in the electron antineutrinos produced by the reactor and the two accessible processes. The first is CE$\nu$NS, $\bar{\nu}_e + \text{N} \rightarrow \bar{\nu}_j + \text{N}$, where $\text{N}$ is a nucleus (Ge in CONUS) and the outgoing neutrino is not measured, so we will sum over the light mass eigenstates $\bar{\nu}_j$. The other is E$\nu$eS, $\bar{\nu}_e + e \rightarrow \bar{\nu}_j + e$. The relevant diagrams are given in figs.~\ref{fig:cevensdiag} and \ref{fig:escattdiag}, which include also the neutrino production vertices. In the unitary (SM) limit, the differential CE$\nu$NS cross section is given by


\begin{figure}
\begin{tikzpicture}
\begin{feynman}
    \vertex  (b) ;
    \vertex [above right=of b](a) {\(e\)};
    \vertex [right=10em of b] (c) ;
    \vertex [above right=of c] (d) {\(\bar{\nu}_j\)};
    \vertex [left=of b] (e);
    \vertex [below left=of e] (i) {\(d\)};
    \vertex [below right=of e] (j) {\(u\)};
    \vertex [below right=of c] (f);
    \vertex [below left=of f] (g) {\(N\)};
    \vertex [below right=of f] (h) {\(N\)};
    \diagram* {
      (a) -- [anti fermion] (b) -- [anti fermion, edge label=\(\bar{\nu}_i\)] (c) -- [anti fermion] (d),
      (e) -- [boson, edge label=\(W\)] (b),
      (c) -- [boson, edge label=\(Z\)] (f),
      (g) -- [fermion] (f) -- [fermion] (h),
      (i) -- [fermion] (e) -- [fermion] (j),
    };
    \vertex [above=0.5em of b] {\(K_{e i}\)};
    \vertex [above=0.5em of c] {\(P_{ij}\)};
  \end{feynman} 
  \draw[gray!25, dashed]
  (2, -2.2) -- (2, 2.2);
  \node at (-0.5, 2) {Production ($\beta^-$ decay)};
  \node at (4.5, 2) {Detection ($\bar{\nu} + N \to \bar{\nu} + N$)};
      \end{tikzpicture}
  \caption{Feynman diagram of CE$\nu$NS. Modifications due to lepton non-unitarity introduce corrections at the neutrino interaction vertices in both the production (left-hand side) and the detection (right-hand side of the diagram).}
  \label{fig:cevensdiag}
  \end{figure}
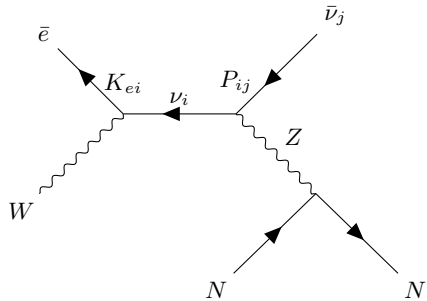


\begin{figure}[t!]
  \centering

  \begin{minipage}{0.45\linewidth}
    \centering
    \begin{tikzpicture}[scale=0.8]
      \begin{feynman}
        \vertex (b) ;
        \vertex [above right=of b] (a) {\(e\)};
        \vertex [left=of b] (e);
        \vertex [below left=of e] (i) {\(d\)};
        \vertex [below right=of e] (j) {\(u\)};

        \vertex [right=6em of b] (c) ;
        \vertex [above right=of c] (d) {\(\bar{\nu}_j\)};

        \vertex [below right=of c] (f);
        \vertex [below left=of f] (g) {\(e\)};
        \vertex [below right=of f] (h) {\(e\)};

        \diagram*{
          (a) -- [anti fermion] (b) -- [anti fermion, edge label=\(\bar{\nu}_i\)] (c) -- [anti fermion] (d),
          (e) -- [boson, edge label=\(W\)] (b),
          (i) -- [fermion] (e) -- [fermion] (j),

          (c) -- [boson, edge label=\(Z\)] (f),
          (g) -- [fermion] (f) -- [fermion] (h),
        };

        \vertex [above=0.5em of b] {\(K_{e i}\)};
        \vertex [above=0.5em of c] {\(P_{ij}\)};
      \end{feynman}

      \draw[gray!25, dashed] (1.85, -2.2) -- (1.85, 2.2);
      \node[font=\footnotesize] at (-0.5, 2.3) {Production ($\beta^-$ decay)};
      \node[font=\footnotesize] at (3.3, 2.3)
{\shortstack{NC Detection\\($\bar{\nu}+e\to\bar{\nu}+e$)}};
    \end{tikzpicture}
  \end{minipage}
  \(\,+\,\)
  \begin{minipage}{0.45\linewidth}
    \centering
    \begin{tikzpicture}
      \begin{feynman}
        \vertex (b) ;
        \vertex [above right=of b] (a) {\(e\)};
        \vertex [left=of b] (e);
        \vertex [below left=of e] (i) {\(d\)};
        \vertex [below right=of e] (j) {\(u\)};

        \vertex [right=6em of b] (c) ;
        \vertex [below left=of c] (d) {\(e\)};

        \vertex [right=of c] (f);
        \vertex [above right=of f] (g) {\(\bar{\nu}_j\)};
        \vertex [below right=of f] (h) {\(e\)};

        \diagram*{
          (a) -- [anti fermion] (b) -- [anti fermion, edge label=\(\bar{\nu}_i\)] (c) -- [anti fermion] (d),
          (e) -- [boson, edge label=\(W\)] (b),
          (i) -- [fermion] (e) -- [fermion] (j),

          (c) -- [boson, edge label=\(W\)] (f),
          (g) -- [fermion] (f) -- [fermion] (h),
        };

        \vertex [above=0.5em of b] {\(K_{e i}\)};
        \vertex [above=0.5em of c] {\(K^{*}_{e i}\)};
        \vertex [right=0.5em of f] {\(K_{e j}\)};
      \end{feynman}

      \draw[gray!25, dashed] (0.15, -2) -- (2, 2);
      \node[font=\footnotesize] at (-0.5, 2.0) {Production ($\beta^-$ decay)};
\node[font=\footnotesize] at (4.3, 2.0)
{\shortstack{CC Detection\\($\bar{\nu}+e\to\bar{\nu}+e$)}};
    \end{tikzpicture}
  \end{minipage}

  \caption{Feynman diagrams for E$\nu$eS mediated by the neutral current (NC, left) or charged current (CC, right). Non-unitarity affects both the production (left-hand side of each diagram) and the detection (right-hand side of each diagram).}
  \label{fig:escattdiag}
\end{figure}


\begin{equation}
    \frac{d\sigma}{dT_A}(T_A, E_\nu) = \frac{G_F^2 }{4\pi} m_A Q_W^2
    \left(1 - \frac{m_A T_A}{2E_\nu^2} \right) |F(T_A)|^2, \quad \text{ with }
    Q_W = \left[(1 - 4\sin^2\theta_W)Z - N \right].
\end{equation}
Here, $m_A$ is the nucleus mass and $Q_W$ is the weak nuclear charge, which is defined by the number of protons $Z$ and neutrons $N$, respectively. For the nuclear form factor, the Helm parameterization is used~\cite{Helm:1956zz}:

\begin{equation}
    F(T_A) = \frac{3j_1(q(T_A)R_1)}{q(T_A)R_1} 
    \exp\left[-\frac{(q(T_A)s)^2}{2} \right],
\end{equation}
with the spherical Bessel function $j_1$, the momentum transfer $q^2 = 2m_A T_A$, the nuclear skin thickness $s \simeq 1$ fm, $R_1 = \sqrt{R^2 - 5s^2}$ and $R \simeq 1.2A^{1/3}$ fm. Due to small momentum transfers at a reactor site, the impact of the nuclear form factor is negligible, i.e.\ $F \to 1$. However, with increasing precision on the SM signal - as investigated in this work - this quantity becomes more relevant.

The target recoil energies $T_x$ for $x = \{e, A\}$ depend on the scattering angle $\theta$ (lab frame) and is given by

\begin{equation}\label{eq:recoil_energy}
    T_x = \frac{2m_x E_\nu^2 \cos^2\theta}{(m_x + E_\nu)^2 - E_\nu^2\cos^2\theta}
    \overset{\theta \to 0}{\longrightarrow} \frac{2E_\nu^2}{m_x + 2E_\nu}\, ,
\end{equation}
where the last step defines the maximal nuclear recoil $T_x^{\text{max}}$.

Compared to the SM, each vertex is modified in the presence of non-unitarity. As such, the probability factor attached to the diagram in \ref{fig:cevensdiag} is given by 

\begin{align}
    \mathcal{P} = N_{ei} N^*_{\alpha i} N_{\alpha j} N^*_{ek} N_{\beta k} N^*_{\beta j} = (N N^\dagger N N^\dagger N N^\dagger)_{ee} \, .
\end{align}
By taking into account the redefinition of the Fermi constant, performing the seesaw expansion and keeping terms up to order $O(\varepsilon^2)$ we find the expected number of CE$\nu$NS events compared to the SM case to be

\begin{equation}
\label{eq:Neventshadron}
    \colorbox{lightblue}{$\displaystyle \left(\frac{\mathcal{N}_\text{NU}}{\mathcal{N}_\text{SM}}\right)^\text{CE$\nu$NS} = \mathcal{P} \frac{G_F^2}{G_\mu^2} = \frac{(N N^\dagger N N^\dagger N N^\dagger)_{ee}}{ (N N^\dagger)_{ee} (N N^\dagger)_{\mu \mu}} \approx 2 \alpha_{11}^2- \alpha_{22}^2$} \, .
\end{equation}
This ratio is equal to $1$ in the unitary limit, where $\alpha_{11} = \alpha_{22} = 1$.


\begin{figure}[t]
\centering
    \includegraphics[width=0.49\textwidth]{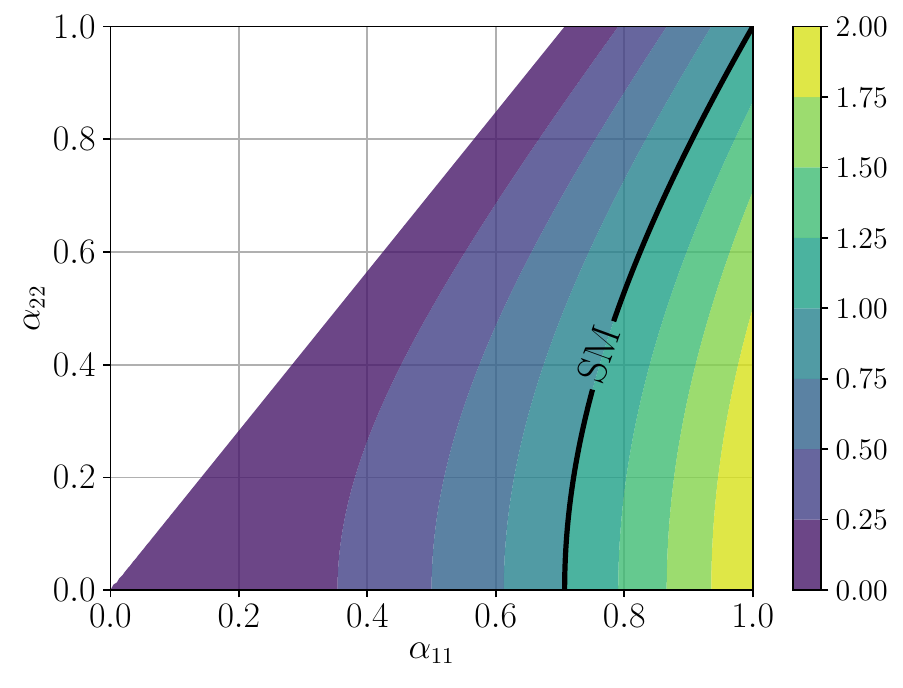}\hfill
    \includegraphics[width=0.50\textwidth]{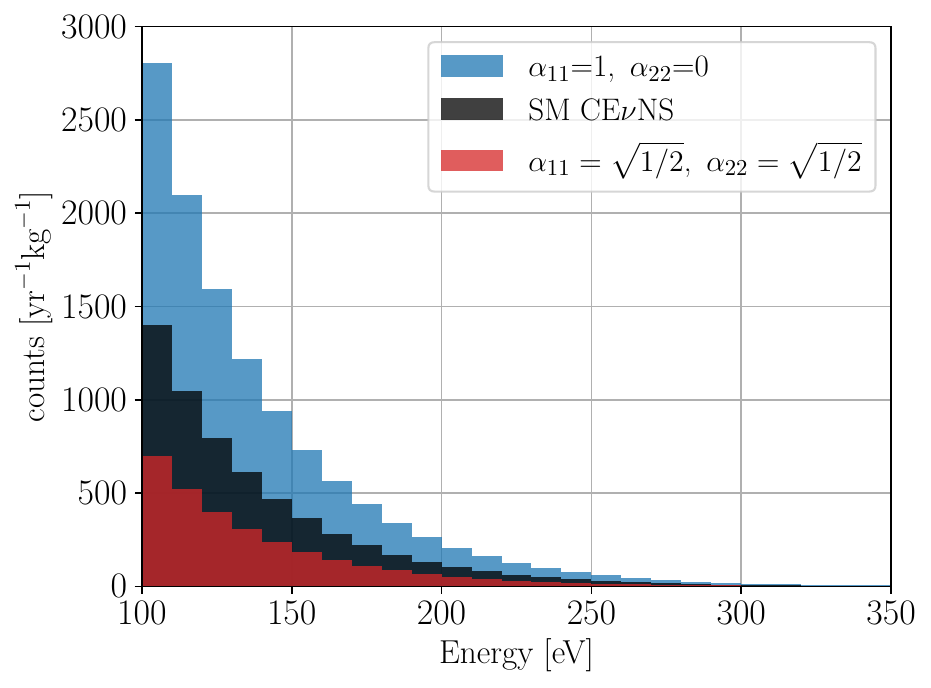}
    \caption{Left: Prefactor $(2\alpha_{11}^{2} - \alpha_{22}^{2})$ present in Eqs.\ \eqref{eq:Neventshadron} and \eqref{eq:Neventslepton}. Depending on the parameter configuration, the expected CE$\nu$NS / E$\nu$eS signal can be smaller or larger than the SM expectation indicated by the black line. Right: Exemplary CE$\nu$NS spectra for the given alpha combinations, which are chosen to receive a bisection and doubling of the events. A flux of $\phi\sim 1.5\cdot10^{13}$/cm$^{2}$/s is assumed ($L=20$\,m, $P_{\mathrm{th}}=3.5$\,GW).}
\label{fig:non_unitarity_characteristics} 
\end{figure}


On the other hand, the differential cross section of the elastic neutrino scattering on electrons in the unitary limit is given by

\begin{equation}
    \frac{d\sigma}{dT_e}(T_e, E_\nu) = \frac{G_F^2 m_e}{2\pi} 
    \left[ (g_V + g_A)^2 + (g_V - g_A)^2 \left(1 - \frac{T_e}{E_\nu}\right)^2 
    + \left(g_A^2 - g_V^2\right)\frac{m_e T_e}{E_\nu^2} \right],
\end{equation}
with $g_V = -\frac{1}{2} + 2\sin^2\theta_W$ and $g_A = -\frac{1}{2}$ for electron antineutrinos.

Now we need to compute the probability factors associated with the neutral and charged current as well as their interference. In general, they are different in the presence of non-unitarity. Therefore, the dependence of the differential cross section on the recoil energy of the final electron will change with respect to the SM. However, this deviation of the shape of the spectrum is not only very hard to measure, but also 'theory suppressed' as we will see. Indeed, the probability factors are given by

\begin{align}
    \mathcal{P}_{NC} & = N_{ei} N^*_{\alpha i} N_{\alpha j} N^*_{ek} N_{\beta k} N^*_{\beta j} = (N N^\dagger N N^\dagger N N^\dagger)_{ee}\, ,  \\
    \mathcal{P}_{CC} & = N_{ei} N^*_{e i} N_{e j} N^*_{ek} N_{e k} N^*_{e j} = (N N^\dagger)_{ee}^3\, ,\\
    \mathcal{P}_{int} & = N_{ei} N^*_{\alpha i} N_{\alpha j}N^*_{ek} N_{e k} N^*_{e j} = (N N^\dagger)_{ee} (N N^\dagger N N^\dagger)_{ee}\, ,
\end{align}
which at order $O(\varepsilon^2)$ in the seesaw expansion become

\begin{equation}
    \mathcal{P}_{NC} \approx \mathcal{P}_{CC} \approx \mathcal{P}_{int} \approx \alpha_{11}^6\, .
\end{equation}
As a consequence

\begin{equation}
\label{eq:Neventslepton}
    \colorbox{lightblue}{$\displaystyle\left(\frac{\mathcal{N}_\text{NU}}{\mathcal{N}_\text{SM}}\right)^\text{E$\nu$eS} = \alpha_{11}^6 \frac{G_F^2}{G_\mu^2} = \frac{\alpha_{11}^6}{ (N N^\dagger)_{ee} (N N^\dagger)_{\mu \mu}} \approx 2 \alpha_{11}^2- \alpha_{22}^2 = \left(\frac{\mathcal{N}_\text{NU}}{\mathcal{N}_\text{SM}}\right)^\text{CE$\nu$NS}$}\, , 
\end{equation}
which is the same modification as in the CE$\nu$NS case.



\begin{figure}[t]
    \centering
    \includegraphics[width=0.49\textwidth]{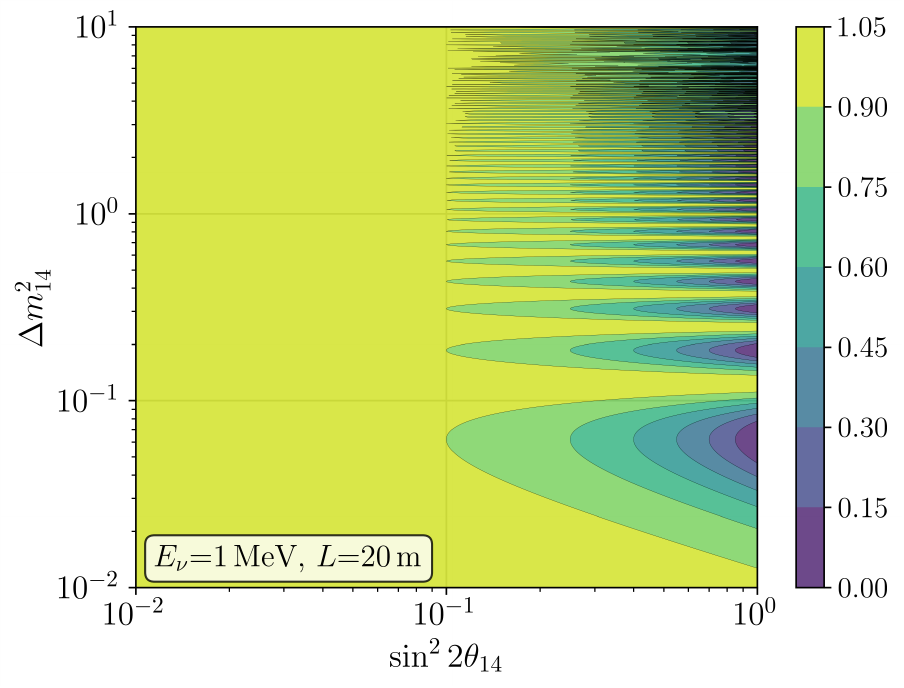}\hfill
    \includegraphics[width=0.5\textwidth]{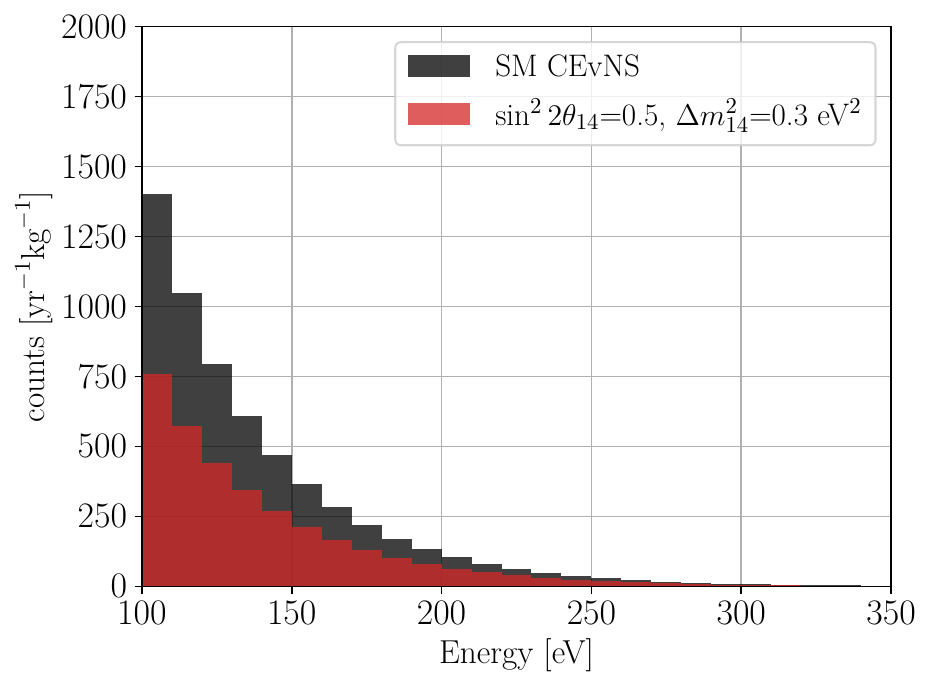}
    \caption{Left: Oscillation probability of Eqs.\ \eqref{eq:oscillation_probability_cenns} and \eqref{eq:oscillation_probability_eves}  present in the CE$\nu$NS and E$\nu$eS cross section for fixed $L/E_{\nu}$. Right: Expected CE$\nu$NS spectrum of a light sterile neutrino for the given parameter combination compared to the SM case. Again, a flux of $\phi\sim 1.5\cdot10^{13}$/cm$^{2}$/s is assumed ($L=20$\,m, $P_{\mathrm{th}}=3.5$\,GW).}
\label{fig:light_sterile_characteristics} 
\end{figure}


\subsection{Light sterile neutrinos}

If instead the gauge singlet states are light enough to be produced and propagate, the full matrix $K$ will appear in the description of the neutrino propagation, instead of only the sub-block $N$. It is also easy to see that, in this case, $G_F$ does not get redefined, since the light sterile states leave the experiment undetected. For concreteness, we focus on the simple $3+1$ case, where we consider the addition of a $4th$ gauge singlet to the $3$ massive active neutrinos. In doing so, we remain agnostic regarding the neutrino mass generation mechanism. In this scenario, the probability factor for the CE$\nu$NS process including oscillation effects is given by 

\begin{align}
    \mathcal{P} = K_{ei} e^{-i E_i t} K^*_{\alpha i} K_{\alpha j} K^*_{ek} e^{i E_k t} K_{\beta k} K^*_{\beta j} = (K e^{-i E t} K^\dagger K e^{i E t} K^\dagger)_{ee}\, ,
\end{align}
where we used $K K^\dagger = 1$. In the basis where the charged lepton mass matrix is diagonal, $K$ is simply the upper $3\times 4$ rectangular block of the full $4\times 4$ mixing matrix $U$. In general, $U$ is parametrized by $6$ angles and $6$ CP-violating phases, making the above expression complicated and not enlightening. Since the mixing between the sterile neutrino and the active states is necessarily small, we can however expand in powers of the small mixing angles $\theta_{4i}$ and identify the rest with the standard mixing angles. Moreover, given the short baseline and the energy of reactor antineutrinos, we can neglect standard oscillation effects, i.e.\ $\Delta E_{ij} t \approx 0$ for $i, j \in \{1,2,3\}$, but we assume that $\Delta E_{4i} t$ is within experimental reach, i.e.\ oscillations from active to sterile states can occur. In this way, we recover the standard electron-electron oscillation survival probability in the presence of a light sterile neutrino at short distance, which only depends on one mixing angle and one mass squared difference:

\begin{equation}
    \mathcal{P} \approx 1- \sin^2 2\theta_{14} \sin^2 \left(\frac{L \Delta m^2_{41}}{4 E_\nu}\right)\, ,
\end{equation}
with the mixing angle $\sin^{2}\theta_{14}$, experimental baseline $L$ and the mass-squared difference $\Delta m^{2}_{41}$. Thus, we obtain for CE$\nu$NS

\begin{equation}\label{eq:oscillation_probability_cenns}
    \colorbox{lightblue}{$\left(\displaystyle\frac{\mathcal{N}_\text{light}}{\mathcal{N}_\text{SM}}\right)^\text{CE$\nu$NS} = 1- \sin^2 2\theta_{14} \sin^2 \left(\frac{L \Delta m^2_{41}}{4 E_\nu}\right)$}\, .
\end{equation}

Similarly, for E$\nu$eS we compute the neutral, charged and interference prefactors being
\begin{align}
    \mathcal{P}_{NC} & = (K e^{-i E t} K^\dagger K e^{i E t} K^\dagger)_{ee} \approx 1- \sin^2 2\theta_{14} \sin^2 \left(\frac{L \Delta m^2_{41}}{4 E_\nu}\right)\, , \\
    \mathcal{P}_{CC} & = K_{ei} e^{-i E_i t} K^*_{ei} K_{ej} K^*_{ej} K^*_{ek} e^{i E_k t} K_{ek} = \left|\left(K e^{-i Et} K^\dagger\right)_{ee}\right|^2\, , \\
    \mathcal{P}_{int} & = K_{ei} e^{-i E_i t} K_{ki}^* K_{kj} K_{el}^* e^{i E_l t} K_{el} K_{ej}^* = \left(K e^{-i E t} K^\dagger K K^\dagger \right)_{ee} \left(K e^{i E t} K^\dagger\right)_{ee} = \mathcal{P}_{CC}\, .
\end{align}

\noindent We can simplify $\mathcal{P}_{CC}$ by again neglecting the standard oscillations and expanding in powers of the small quantities $\theta_{4i}$ and find

\begin{equation}
     \colorbox{lightblue}{$\mathcal{P} \equiv \mathcal{P}_{NC} \approx \mathcal{P}_{CC} \approx \mathcal{P}_{int} \approx 1- \sin^2 2\theta_{14} \sin^2 \left(\frac{L \Delta m^2_{41}}{4 E_\nu}\right)$}\, ,
\end{equation}
which implies 
\begin{equation}\label{eq:oscillation_probability_eves}
    \colorbox{lightblue}{$\displaystyle\left(\frac{\mathcal{N}_\text{light}}{\mathcal{N}_\text{SM}}\right)^\text{E$\nu$eS} = 1- \sin^2 2\theta_{14} \sin^2 \left(\frac{L \Delta m^2_{41}}{4 E_\nu}\right) = \left(\displaystyle\frac{\mathcal{N}_\text{light}}{\mathcal{N}_\text{SM}}\right)^\text{CE$\nu$NS}$}\, .
\end{equation}


\section{Experimental framework and analysis}\label{sec:experiment}


\noindent Here we assume an upscaling of the Germanium-based detector technology used in CONUS+ that has recently achieved the first detection of CE$\nu$NS with reactor antineutrinos~\cite{Ackermann:2025obx}. This technology allows to go to larger detector masses which we scale up to 100\,kg. 
We consider plausible extrapolations of the current Ge-based concept. Since the mass scales with the volume ($M\propto L^3$), a moderate increase in size already yields a substantial gain in mass. Using the current CONUS+ shield design with its capacity of housing $\sim$10\,kg Ge inside a detector chamber of $L\sim0.3$\,m, the $\mathcal{O}(100\,\mathrm{kg})$ mass scale corresponds to an increase in size by a factor of $\sim2.2$, which may be achieved, among other things, by arranging several detectors in one cryostat. Concerning shielding, high-density low-activity materials (e.g.\ clean tungsten, as used within the CONUS+ collaboration) allow for more compact shielding designs than conventional lead-based solutions. Finally, while the feasibility of deploying a given mass--shielding configuration at a baseline of $\sim$20\,m ultimately depends on site-specific infrastructure, safety/licensing, space and statics constraints, the 20\,m distance should be regarded as a representative near-field benchmark. In limiting cases, dividing the target mass into several experimental set-ups, each with its own shielding, could be considered. In particular, set-ups at different distance can have advantages in dedicated physics investigations, as we will comment on below.

In addition to their mass, Germanium detectors have a low energy threshold~\cite{CONUS:2024lnu} for which we also consider conceivable improvements. 
We assume as source nuclear power reactors that have a very intense electron-antineutrino flux. Specifically we assume the experiment to be located at a 20\,m-distance to a typical commercial reactor with a thermal power of 3.5\,GW.
We choose a typical fuel composition of a pressurized water reactor ($^{235}$U, $^{238}$U, $^{239}$Pu, $^{241}$Pu) = (56.1, 7.6, 30.7, 5.6)\ \% and select a data-based reactor antineutrino spectrum. For this, we use the method proposed in \cite{DayaBay:2021dqj} and utilize the provided unfolded spectra of inverse beta decay (IBD) measurements. The high energy part of the spectrum provided by \cite{DayaBay:2022eyy} and calculated spectra for the energy region below the IBD threshold of \cite{Estienne:2019ujo} were added with appropriate normalization. Similar spectra have already been used in previous data analyses~\cite{Ackermann:2025obx,CONUSCollaboration:2024kvo}.
This leads to an electron antineutrino flux of $\sim1.5\cdot10^{13}$/cm$^{2}$/s at the experimental site with underlying uncertainties - among others, of the reactor thermal power and fission fractions. A 3\% level uncertainty can be considered established, but we will also show the impact of an improved uncertainty that could arise from combining all other existing and upcoming reactor experiments as the field moves toward a precision, multi-experiment frontier. The ``optimized'' 0.3\% flux uncertainty should not be interpreted as a near-term improvement in reactor flux calculations, but rather as an effective reduction of the flux normalization uncertainty enabled by combining multiple high-statistics reactor neutrino measurements in global fits.

The cross sections of CE$\nu$NS and E$\nu$eS weighted with the spectral information are obtained via
\begin{align}
    \frac{d\sigma}{dT_{x}}(T_{x}) = \int_{E_{\mathrm{min}}}^{E_{\mathrm{max}}} dE_{\nu} \frac{dN}{dE_{\nu}}(E_{\nu}) \frac{d\sigma}{dT_{x}}(T_{x}, E_{\nu})\, ,
\end{align}
with $T_{\{e^{-},\ N\}}$ being the electron and nuclear recoil, respectively, $E_{\nu}$ the antineutrino energy and $dN/dE_{\nu}$ the reactor spectrum reaching up to $E_{\mathrm{max}}\simeq11$\,MeV. 
The minimal recoil energy is set by Eq.\ \eqref{eq:recoil_energy}.

Further, we assume flat background levels of 10\,cnts/keV/kg/d in the region of interest (ROI) for CE$\nu$NS investigations, i.e.\ below 1\,keV, and background contributions of 0.5\,cnts/keV/kg/d above as reference values. A rise in background events towards lower energies is usually expected and will degrade the sensitivity of configurations with lowest thresholds. However, realistic background models depend on various (site-specific) factors, such as radiopurity of used materials, distance to the reactor and overburden as well as the experimental environment in the reactor building itself. Therefore, our assumptions should be viewed as an effective description of background in the ROI. More details about background events in reactor environments and individual contributions can be found in \cite{Bonet:2021wjw,CONUS:2024vyx}.
To reduce uncertainties on the background, we fit times of operating (ON) and shut down (OFF) of the reactor simultaneously and assume $t_{\mathrm{OFF}} = 0.1\cdot t_{\mathrm{ON}}$.\footnote{Nuclear power reactors typically run for eleven months followed by a month for re-fueling.} Effects of overall improved background levels are discussed below. In this study we assume experimental exposures of (5, 50, 500)\,kg$\cdot$yrs (reactor ON). 5\,kg$\cdot$yrs corresponds to CONUS+ operation which we term as ``now''. 50\,kg$\cdot$yrs is something that can be obtained by operating CONUS+ with the latest upgraded detectors for a few years, which we call ``soon''. 500\,kg$\cdot$yrs corresponds to an upscaling with 100\,kg detector mass operated for 5~years, which we call ``future'', see Tab.~\ref{tab:benchmarks} for further details.

\begin{table}[t]
\centering
\resizebox{\textwidth}{!}{
\begin{tabular}{l c c c c c p{6.5cm}}
\hline\hline
Scenario & exp.\ [kg$\cdot$yr] & $E_{\rm thr}$ [eV] & B [cnts/kg/keV/d] & $\Delta\Phi$ & $\Delta k$ & Notes \\
\hline
now    & 5   & 150 & (10, 0.5) & 3\%  & 2\% & CONUS+ operation scale \\
soon   & 50  & 125 & (10, 0.5) & 3\%  & 2\% & Few-year operation with upgraded detectors \\
future & 500 & 100 & (10, 0.5) & 3\%  & 2\% & $100$ kg $\times$ $5$ yr upscaling \\\hline
optimized & X & X & (1, 0.05) & 0.3\% & 1\% & Same exposure/threshold\\
\hline\hline
\end{tabular}
}
\caption{Benchmark experimental configurations used in this work. The experimental background is divided into two regions: below 1\,keV (first value) and above 1\,keV (second value). The category ``optimized'' refers to improved experimental systematics of the given categories.}
\label{tab:benchmarks}
\end{table}

For the Germanium detectors we assume 100\% detection efficiency in the ROI down to their threshold energies, for which we choose a (``now'', ``soon'', ``future'') value of (150, 125, 100)\,eV. While operating a detector under such circumstances is a challenge in itself, we assume it to be feasible for our study with sufficient long-term stability. Our choice is motivated by current CONUS+ R\&D: a 2.4\,kg Ge prototype with an energy threshold of $\mathcal{O}(100\,{\rm eV})$ has already been achieved under realistic conditions. Since this detector type is only sensitive to ionization, we need to account for the conversion of nuclear recoils $T$ into charge signals $E$.
The relevant quenching factor is so far well described by the widely applied Lindhard model~\cite{Lindhard:1961zz} that has been confirmed to be valid for Germanium semiconductor detector down to the energies of interest. We use for the theory's $k$ parameter (reflecting the ratio between ionization and recoil energy at $~1$\,keV) the measured value of $k=0.162\pm 0.004$~\cite{Bonhomme:2022lcz}. Future measurements may reduce the uncertainty on the $k$ parameter and we will discuss the impact of such improvements.
The conversion from recoil to ionization energy is done by a variable transform
\begin{align}
    \frac{d\sigma}{dE} (E) = \left[ Qf^{-1} (E) + E\, \frac{d\, Qf^{-1}}{dE}  (E) \right] \frac{d\sigma}{dT_{N}} (Qf^{-1}(E) \cdot E)\, ,
\end{align}
with $Qf^{-1}(E)$ being the inverted Lindhard model in terms of ionization (detected) energy $E$.

Furthermore, we assume a connection between the detector's threshold energy and the intrinsic noise of the read-out electronic. In particular, we impose the detector threshold to be three times the FWHM (full width at half maximum) of an artificial test pulse at zero energy $E_{\mathrm{thr}}\sim 3\cdot\mathrm{FWHM}$. 
As a consequence, the detector resolution is described by a Gaussian with an energy-dependent width given by
\begin{align}
    \sigma^{2}(E) = \left( \frac{E_{\mathrm{thr}}}{3\cdot 2.355}\right)^{2} + \epsilon \cdot F \cdot E\, ,
\end{align}
with the energy necessary to create an energy-hole pair in Germanium $\epsilon=2.96$\,eV (at 90\,K) and the so-called Fano factor $F=0.11$. 

Taking all these aspects into account, the expected SM events rate are (16, 9, 5)\,cnts/kg/d for CE$\nu$NS in the region (100, 125, 150)\,eV up to 1\,keV and $\sim1.4$\,cnts/kg/d/keV for E$\nu$eS up to 100\,keV. 
Exemplary CE$\nu$NS spectra for the scenarios under consideration can be found in the right plots of the figures \ref{fig:non_unitarity_characteristics} and \ref{fig:light_sterile_characteristics}, respectively.

Our sensitivity estimates are obtained with a Likelihood function that incorporates both reactor ON and reactor OFF data and pull terms taking into account the experimental uncertainties of the neutrino flux $\Delta\Phi$ and the Lindhard model $\Delta k$,
\begin{align}
    -2\log \mathcal{L} = -2\log \mathcal{L}_{\mathrm{ON}} -2\log \mathcal{L}_{\mathrm{OFF}} + \text{pull terms}\, .
\end{align}
Two model parameters - $(\alpha_{11},\, \alpha_{22})$ in the seesaw and $(\sin^{2}2\theta_{14},\, \Delta m^{2}_{41})$ in the light sterile limit - are fit together with two background normalization parameters $b_{<1\,\mathrm{keV}}$ and $b_{>1\,\mathrm{keV}}$, while 3\% and 1\% uncertainties were assumed for the reactor antineutrino flux $\Delta\Phi$ and quenching given by the Lindhard model $\Delta k$, respectively. In addition, we allow the Weinberg angle $\sin^{2}\theta_{W}$ to vary within current uncertainties at low energy $\sin^{2}\theta_{W}=0.2374 \pm 0.0020$~\cite{AtzoriCorona:2023ktl}.
Additionally, we determine results for a factor 10 improvement in $\Delta \Phi$ and the assumed background levels as well as a factor 2 in $\Delta k$ to quantify the impact of these parameters. In doing so, we underline which experimental parameters are worth improving for future experiments.

We perform a likelihood ratio test and extract limits (at 90\% C.L.) on the parameter space from a $\chi^{2}$-distributed test statistics with two degrees of freedom. 
For the seesaw limit, we incorporate knowledge extracted from oscillation experiments~\cite{Forero:2021azc} by adding an additional two-dimensional pull term to the likelihood function above.
Of course, a proper treatment would require a combined investigation (global fit) of both datasets including underlying experiment-specific systematic effects, which are beyond the scope of this work.
However, in this way we want to underline the potential of future CE$\nu$NS reactor experiments in such ``global fits'' when data from several neutrino experiments are combined.


\section{Results}\label{sec:results}


\subsection{Seesaw limit}

As anticipated from the count rates mentioned in the previous section, our limits are mainly driven by the CE$\nu$NS channel. 
In addition, this has been confirmed by investigating the impact of both interaction channels individually, cf.\ figure \ref{fig:single_channel_alpha_parameter}.
Although there exist already strong constraints from oscillation experiments~\cite{Forero:2021azc}, an improvement for $\alpha_{11}$ is expected due to an appearing factor of 2 in the prefactors of Eqs.\ \eqref{eq:Neventshadron} and \eqref{eq:Neventslepton}.

\begin{figure}[H]
    \centering
    \begin{minipage}{\textwidth}
        \includegraphics[width=0.99\textwidth]{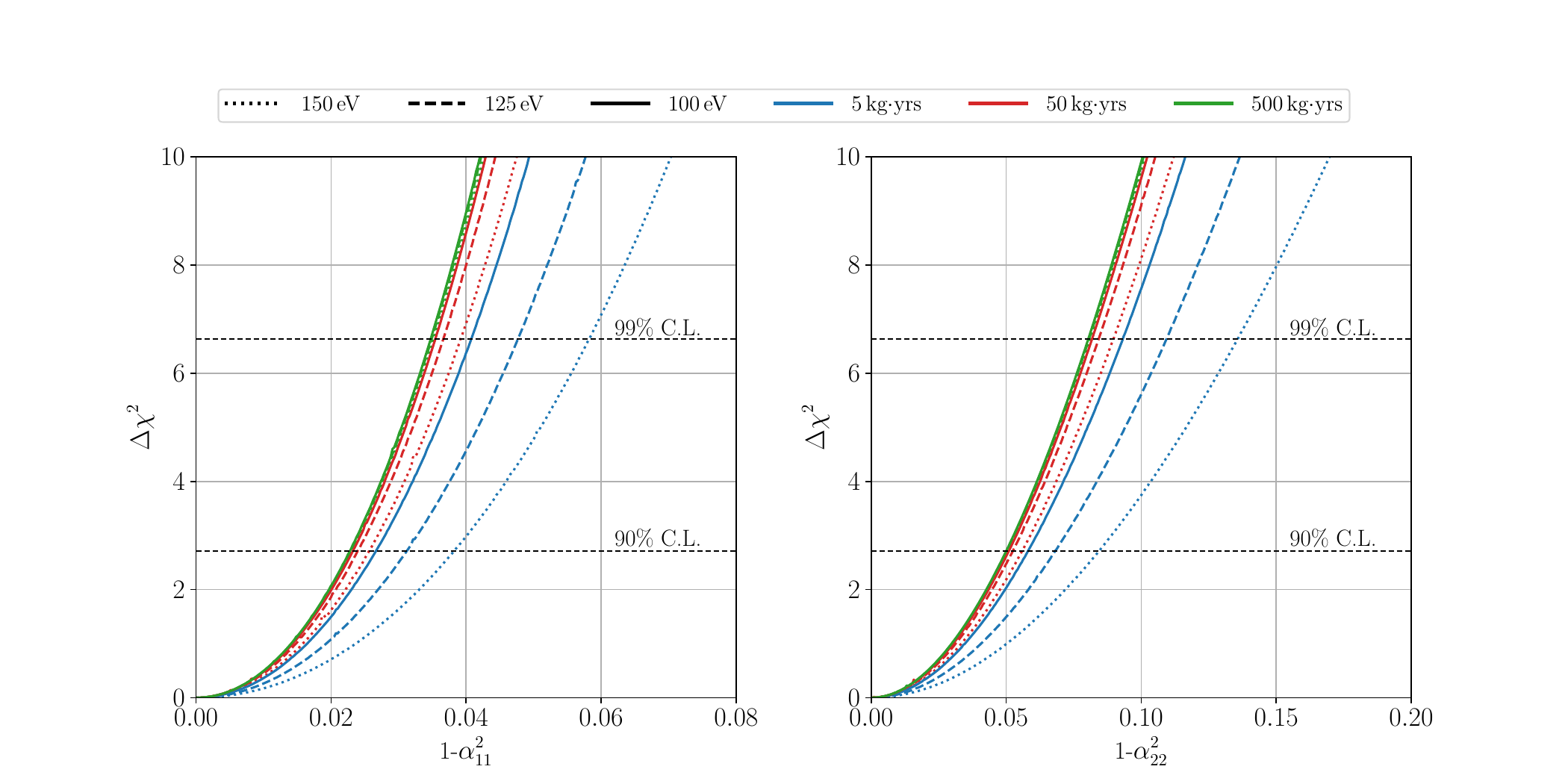}
        \caption{$\Delta \chi^{2}$ profiles of the individual alpha parameters - the other one being fixed to unity - for the assumed experimental configuration. We show three threshold configurations together with three assumptions on the experimental exposure, which are indicated by different colors and line styles, see  Tab.~\ref{tab:benchmarks} for full descriptions. Detailed profiles for the individual thresholds also combined with knowledge from oscillation experiments are illustrated in figure~\ref{fig:alpha_profiles_detailed} in the appendix.}
        \label{fig:alpha_profiles} 
    \end{minipage}

\vspace{2cm}

    \begin{minipage}{\textwidth}
    \centering
    \begin{tabular}{C{2.5cm}C{2.5cm}|C{2cm}C{2cm}|C{2cm}C{2cm}}
    \multicolumn{2}{c|}{} & \multicolumn{2}{c|}{$1-\alpha_{11}^{2}$} & \multicolumn{2}{c}{$1-\alpha_{22}^{2}$} \\ \hline
     &   & & + osci. &  & + osci. \\
    150\,eV  & 5\,kg$\cdot$yr     & 0.039 & 0.022 & 0.086 & 0.008 \\
            & 50\,kg$\cdot$yr    & 0.026 & 0.018 & 0.057 & 0.008 \\
            & 500\,kg$\cdot$yr   & 0.023 & 0.016 & 0.051 & 0.008\\ 
    \hline
    125\,eV  & 5\,kg$\cdot$yr     & 0.032 & 0.020 & 0.069 & 0.008 \\
            & 50\,kg$\cdot$yr    & 0.024 & 0.017 & 0.053 & 0.008 \\
            & 500\,kg$\cdot$yr   & 0.023 & 0.016 & 0.051 & 0.008 \\ 
     \hline
     100\,eV  & 5\,kg$\cdot$yr    & 0.027 & 0.018 & 0.059 & 0.008 \\
            & 50\,kg$\cdot$yr    & 0.024 & 0.016 & 0.052 & 0.008 \\
            & 500\,kg$\cdot$yr   & 0.023 & 0.016 & 0.051 & 0.008 \\
      \hline
    \multicolumn{2}{c|}{oscillations (90\% CL) \cite{Forero:2021azc}} & \multicolumn{2}{c|}{0.061} & \multicolumn{2}{c}{0.01} \\ \hline
    \end{tabular}
    \captionof{table}{Individual $90\%$ C.L.\ limits on the alpha parameters ($1-\alpha_{ii}^{2}$) (the other one fixed to unity) for the assumed experimental configuration. The left columns show the limits for our experimental configuration alone, while right columns indicate limits obtained when knowledge from oscillation experiments is incorporated. The corresponding (oscillation) limits are given in the bottom row.}
    \label{tab:single_alpha_limits}
    \end{minipage}
\end{figure}

\begin{figure}[H]
    \centering
    \begin{minipage}{\textwidth}
        \centering
        \includegraphics[width=0.99\textwidth]{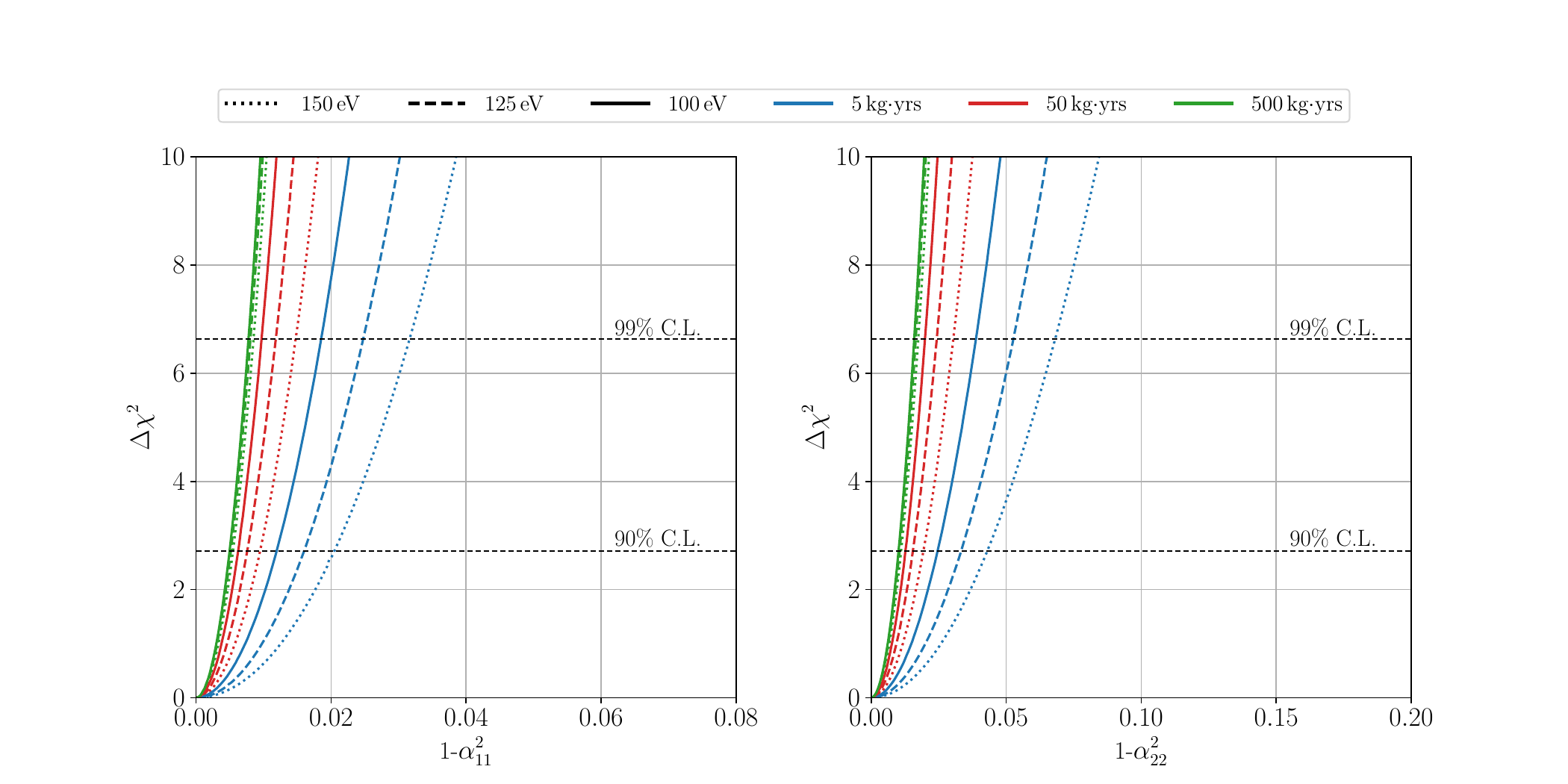}
        \caption{$\Delta \chi^{2}$ profiles of the individual alpha parameters - the other one being fixed to unity - for the optimized experimental configuration. Again, we show three threshold configurations together with three experimental exposures, which are indicated by different colors and line styles. Detailed descriptions of our benchmark configurations are given in Tab.~\ref{tab:benchmarks}.}
        \label{fig:alpha_profiles_opt} 
    \end{minipage}

\vspace{2.5cm}
    \begin{minipage}{\textwidth}
        \centering
        \begin{tabular}{C{2.5cm}C{2.5cm}|C{2cm}C{2cm}|C{2cm}C{2cm}}
        \multicolumn{2}{c|}{} & \multicolumn{2}{c|}{$1-\alpha_{11}^{2}$} & \multicolumn{2}{c}{$1-\alpha_{22}^{2}$} \\ \hline
         &   & & + osci. &  & + osci. \\
        150\,eV  & 5\,kg$\cdot$yr     & 0.021 & 0.015 & 0.044 & 0.008 \\
                & 50\,kg$\cdot$yr    & 0.010 & 0.008 & 0.020 & 0.007 \\
                & 500\,kg$\cdot$yr   & 0.006 & 0.005 & 0.011 & 0.006\\ 
        \hline
        125\,eV  & 5\,kg$\cdot$yr     & 0.016 & 0.013 & 0.034 & 0.008 \\
                & 50\,kg$\cdot$yr    & 0.008 & 0.007 & 0.016 & 0.007 \\
                & 500\,kg$\cdot$yr   & 0.005 & 0.005 & 0.011 & 0.006 \\ 
         \hline
         100\,eV  & 5\,kg$\cdot$yr    & 0.012 & 0.010 & 0.025 & 0.008 \\
                & 50\,kg$\cdot$yr    & 0.006 & 0.006 & 0.013 & 0.007 \\
                & 500\,kg$\cdot$yr   & 0.005 & 0.005 & 0.010 & 0.006 \\
          \hline
        \multicolumn{2}{c|}{oscillations (90\% CL) \cite{Forero:2021azc}} & \multicolumn{2}{c|}{0.061} & \multicolumn{2}{c}{0.01} \\ \hline
        \end{tabular}
        \captionof{table}{Individual $90\%$ C.L.\ limits on the alpha parameters ($1-\alpha_{ii}^{2}$) (the other one fixed to unity) for the optimized experimental configuration. The left columns shows the limits from our experimental configuration alone, while right columns indicate limits obtained when knowledge from oscillation experiments is incorporated. The corresponding (oscillation) limits are given in the bottom row.}
        \label{tab:single_alpha_limits_opt}
    \end{minipage}
   
\end{figure}

At first we assess the sensitivity to the individual non-unitarity parameters by fixing the other one to unity. 
Our results are summarized in figure \ref{fig:alpha_profiles} and table \ref{tab:single_alpha_limits} for our reference configuration and figure \ref{fig:alpha_profiles_opt} and table \ref{tab:single_alpha_limits_opt} for optimized experimental features, respectively.
Detailed $\Delta\chi^{2}$ profiles for the first, also when combined with oscillation data, can be found in the appendix, cf.\ figure \ref{fig:alpha_profiles_detailed}.
We note that limits improve with better detection thresholds and increased exposure.
However, advances are stronger for the transition from recent to soon available thresholds and the first increase in exposure.
For the highest assumed exposure, there is no clear improvement regardless of the chosen detector threshold, cf.\ figure \ref{fig:alpha_profiles}.
This indicates that the assumed systematic uncertainties, i.e.\ the antineutrino flux and signal quenching, become dominant. 
A general improvement is obtained when our reactor-only analysis is combined with information from oscillation experiments.
While for $\alpha_{22}$ existing bounds from oscillation experiments are already quite strong, limits on $\alpha_{11}$ improve in a combined analysis, cf.\ table \ref{tab:single_alpha_limits}, clearly underlining the importance of combined approaches in the future.

Looking at the configuration with reduced uncertainties and lower background level (a factor 10 for flux and background and a factor 2 for quenching), we obtain limits a factor $>2$ better than for the previous case, cf.\ table \ref{tab:single_alpha_limits_opt}.
Especially the larger exposures benefit from this optimization and yield increasingly better constraints. 
When combined with knowledge from oscillation, limits further improve with lower detector threshold and larger exposure.
The obtained limits can be converted into an approximate mass scale where the connected new particles are expected to appear, cf.\ Eq.\ \eqref{eq:limit_to_mass}, assuming a low-scale seesaw and $\mathcal{O}(1)$ parameters. 
In doing so, we could soon (50\,kg$\cdot$yr exposure and 125\,eV-threshold) constrain new physics to lie above $\sim 1100$\,GeV and $\sim 760$\,GeV for $\alpha_{11}$ and $\alpha_{22}$, respectively.
An improved setup could lift these ``bounds'' up to $\sim 1900$\,GeV ($\alpha_{11}$) and $\sim 1400$\,GeV ($\alpha_{22}$).\footnote{Here, we chose the intermediate values for exposure and detector threshold to show the experimental potential that is going to be available in the near future.}
%




\begin{figure}[H]
\centering
    \includegraphics[width=0.49\textwidth]{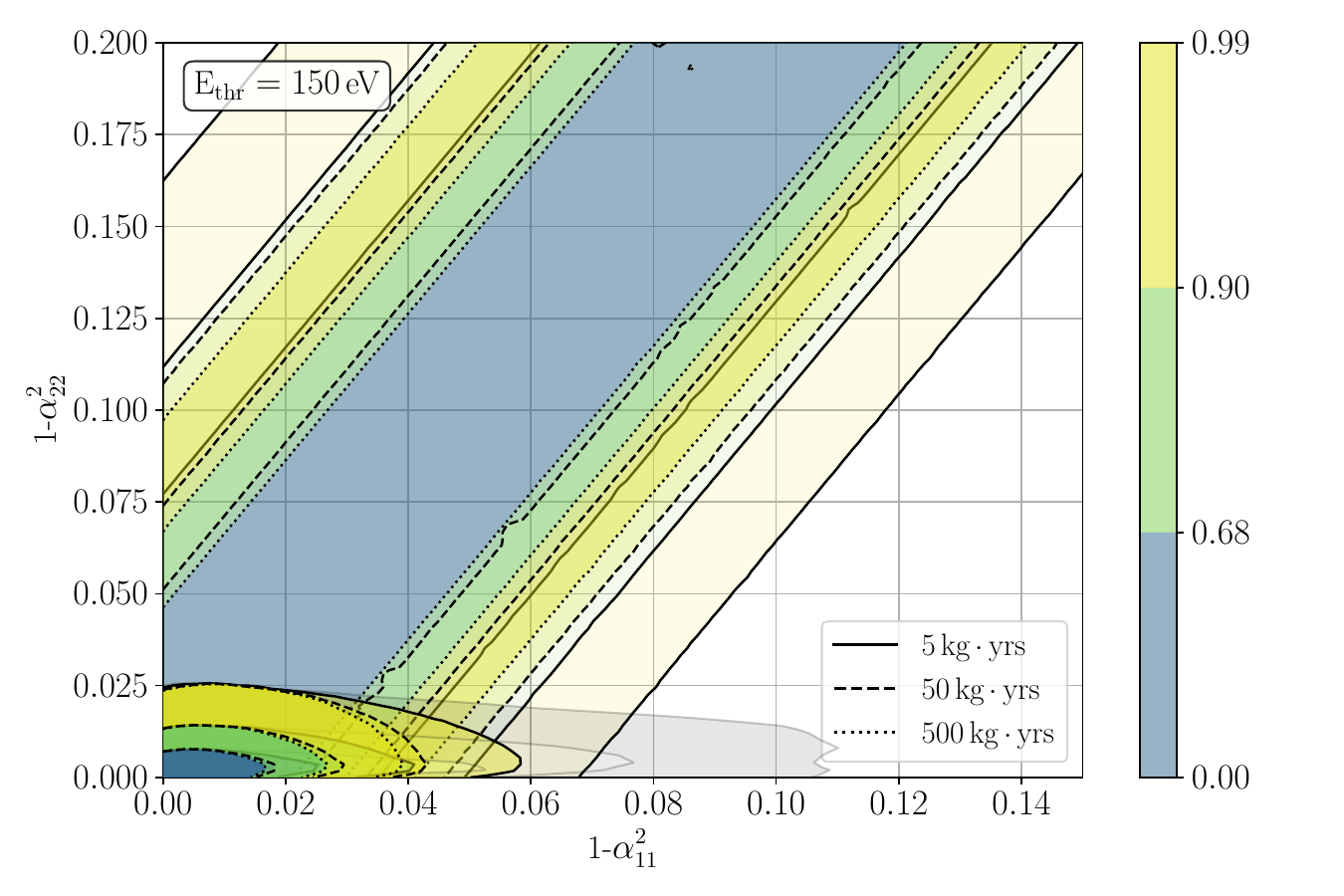}
    \hfill
    \includegraphics[width=0.49\textwidth]{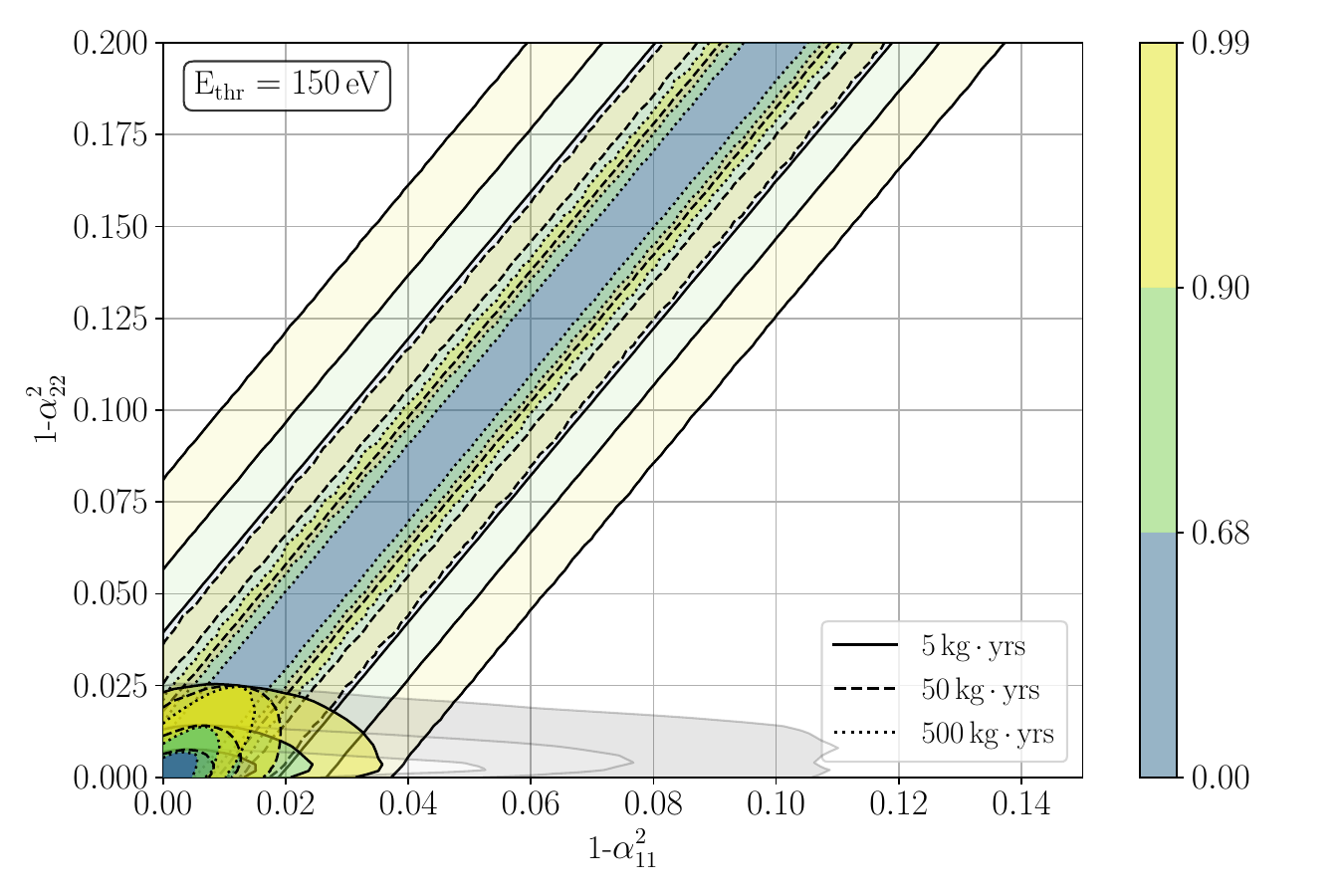}\\
    \vspace{1cm}
    \includegraphics[width=0.49\textwidth]{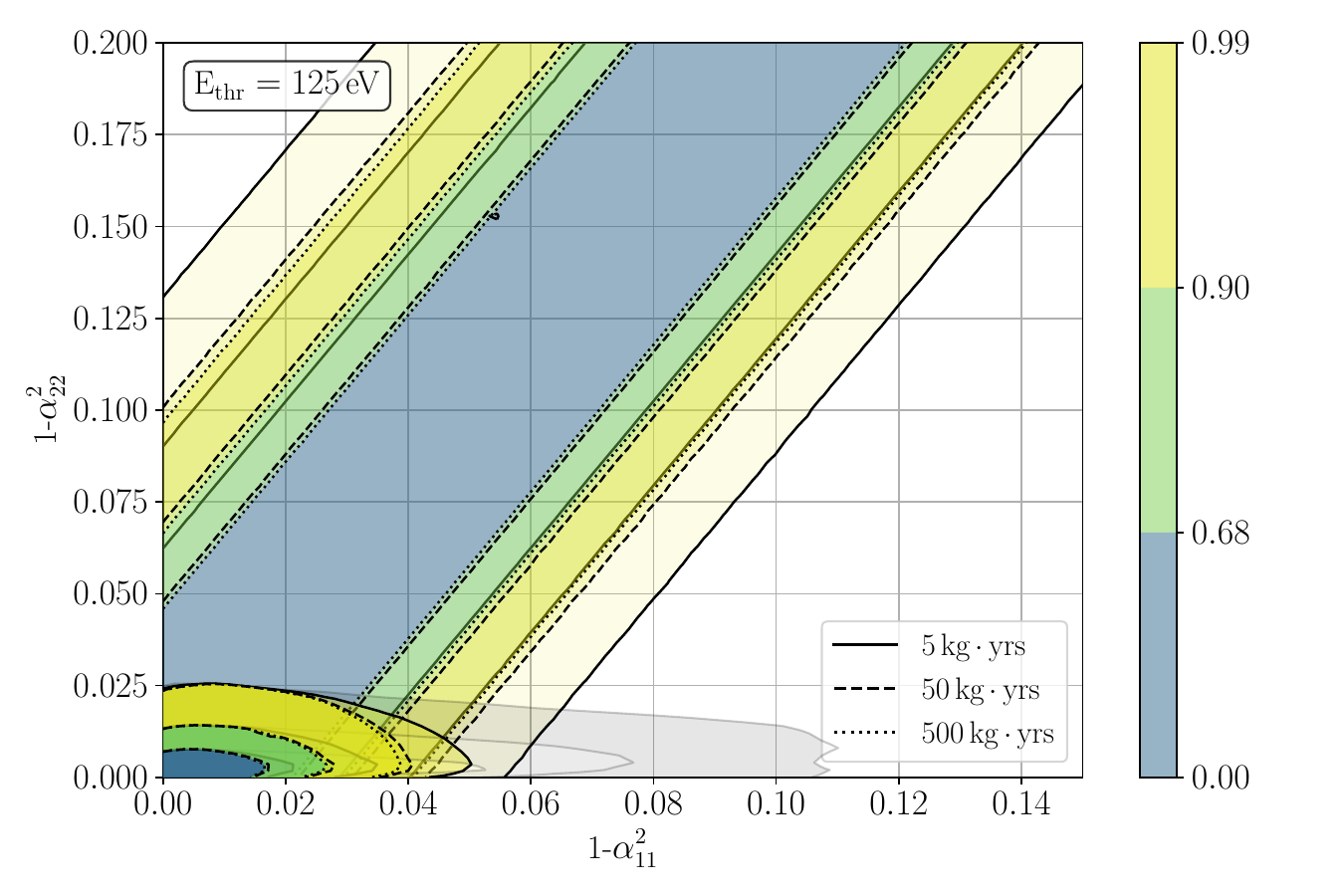}
    \hfill
    \includegraphics[width=0.49\textwidth]{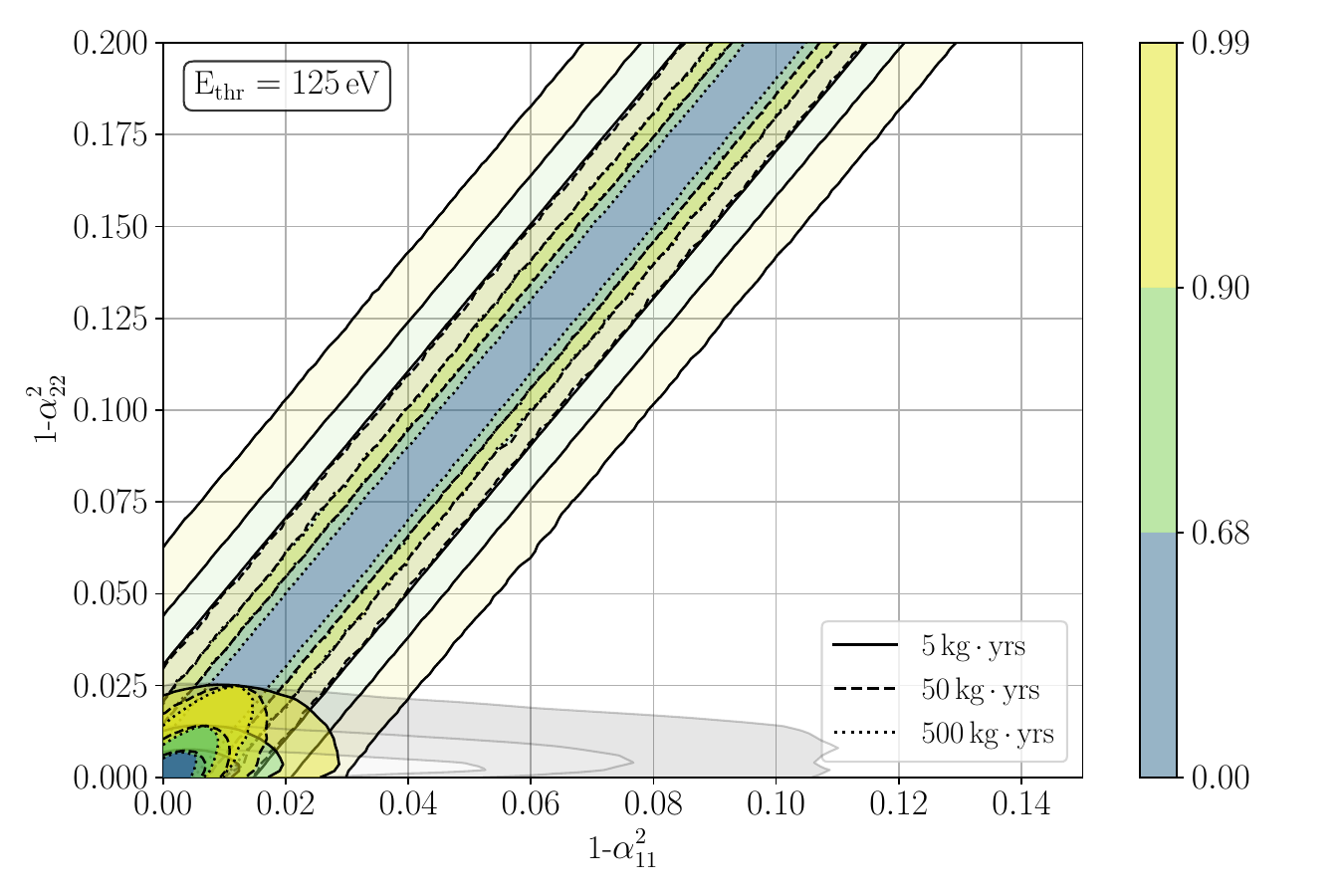}\\
    \vspace{1cm}
    \includegraphics[width=0.49\textwidth]{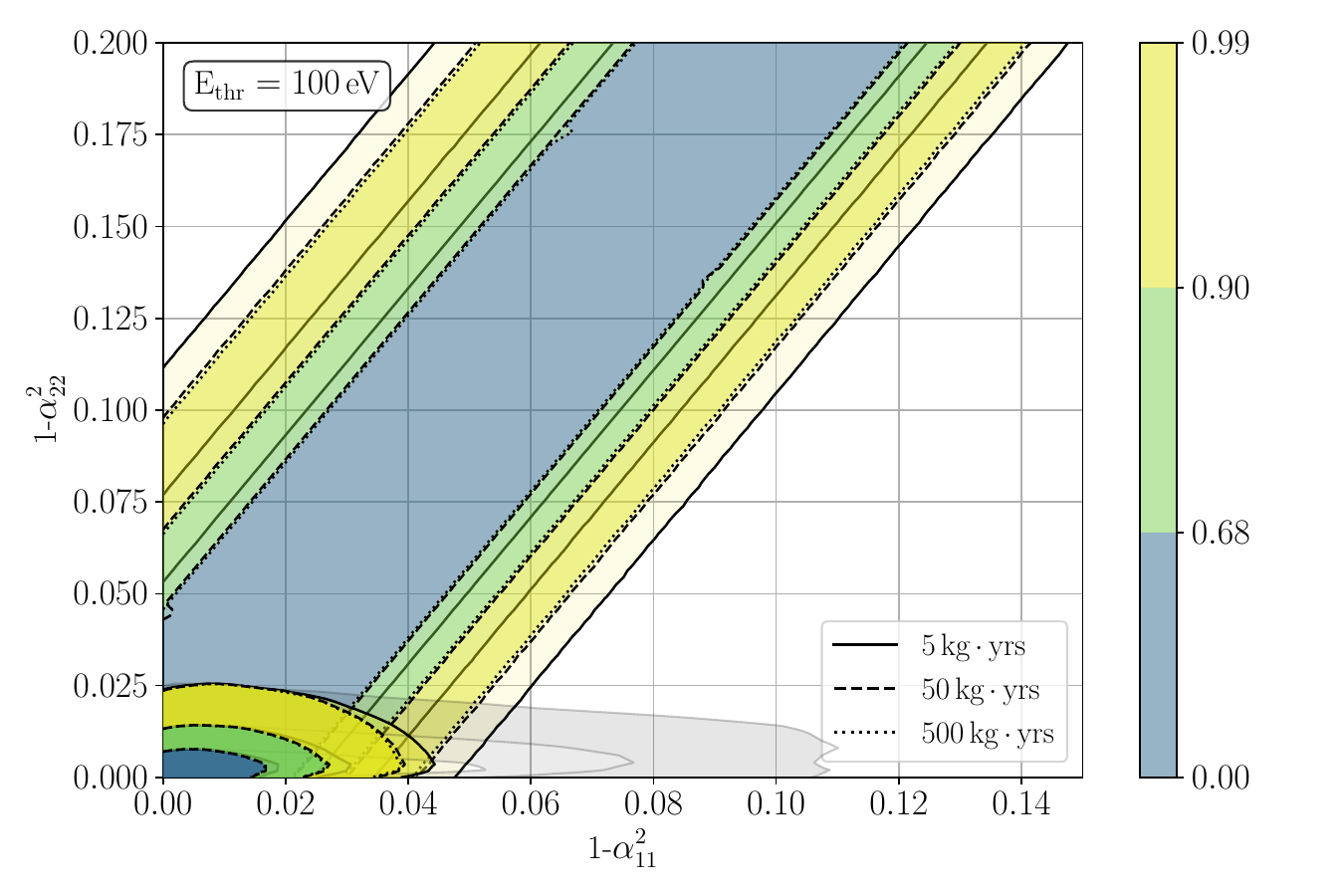}
    \hfill
    \includegraphics[width=0.49\textwidth]{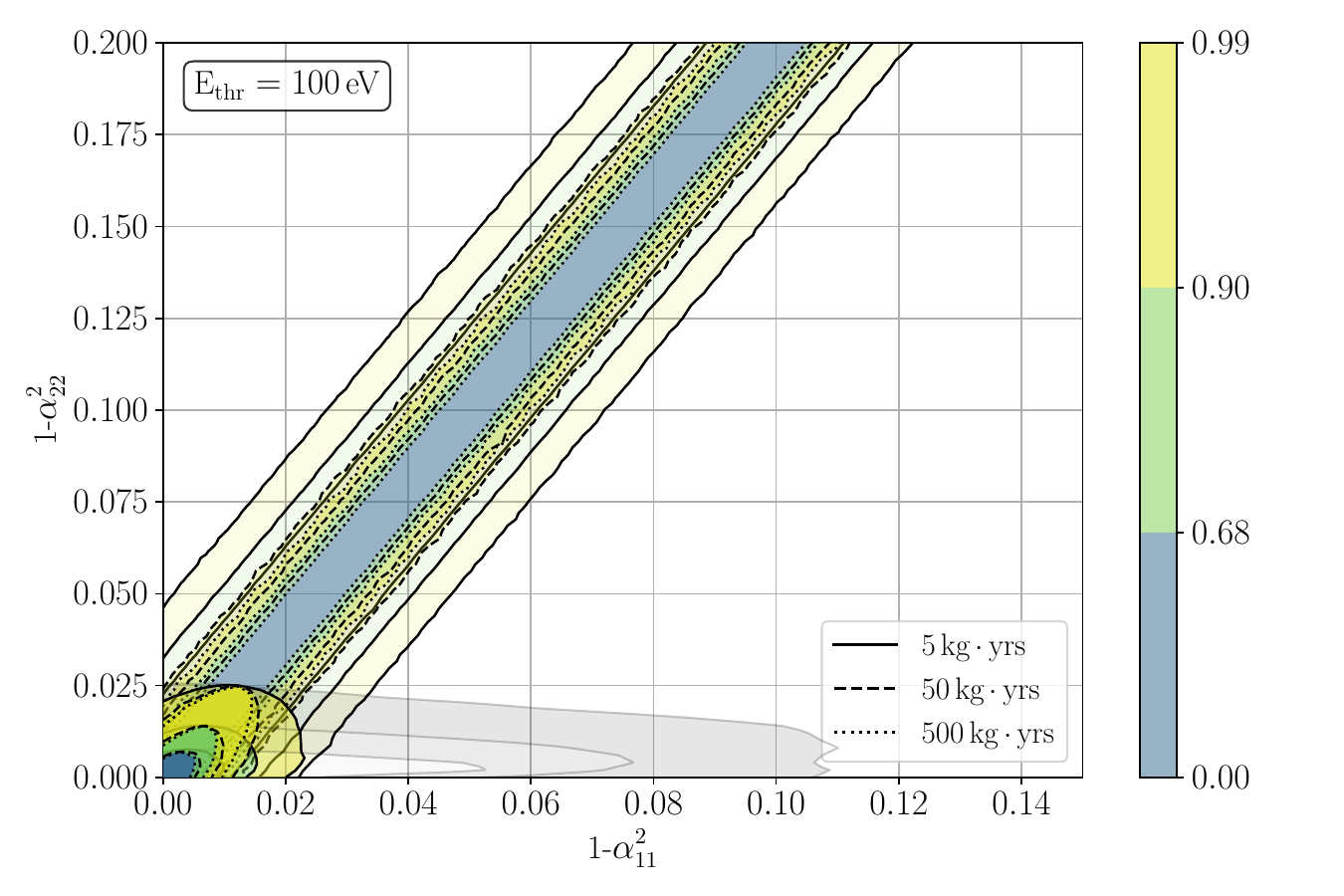}
    \caption{Allowed regions of the alpha parameters for three threshold values and three exposures for our detector. The transparent contours show the limits of the assumed experimental configuration alone, while the non-transparent regions indicate parameter space still allowed when combined with knowledge from oscillation experiments. Gray contours show limits from oscillation experiments alone. Left: Results for our reference setup.  Right: Results for improved knowledge on quenching (factor 2), background (factor 10) and reactor antineutrino flux (factor 10). Descriptions of all configurations are given in Tab.~\ref{tab:benchmarks}.}
\label{fig:alpha_parameter_space} 
\end{figure}

An optimistic scenario with a 100\,eV-threshold and 500\,kg$\cdot$yr combined with oscillations and the optimized experimental setup would constrain new physics up to 2500\,GeV.

The results of the full investigation are given in figure \ref{fig:alpha_parameter_space}.
As for the case of the single parameter investigation, we see the systematic uncertainties become limiting. 
For a threshold of 150\,eV, there is a clear improvement between the exposures 5\,kg$\cdot$yrs and 50\,kg$\cdot$yrs. 
The soon available threshold value, i.e.\ 125\,eV, already shows only a minor improvement between 50\,kg$\cdot$yrs and 500\,kg$\cdot$yrs, while for a 100\,eV-threshold there is almost no increase in sensitivity.
With the help of table \ref{tab:single_alpha_limits}, it is possible to identify experimental configuration of almost similar sensitivity: 500\,kg$\cdot$yrs of exposure with a 150\,eV-threshold is complementary to 50\,kg$\cdot$yrs exposure with a 100\,eV-threshold. 
Such information are valuable and further development on the experimental site might decide which path to follow in the future.


 \begin{figure}[t]
        \centering
        \includegraphics[width=0.99\textwidth]{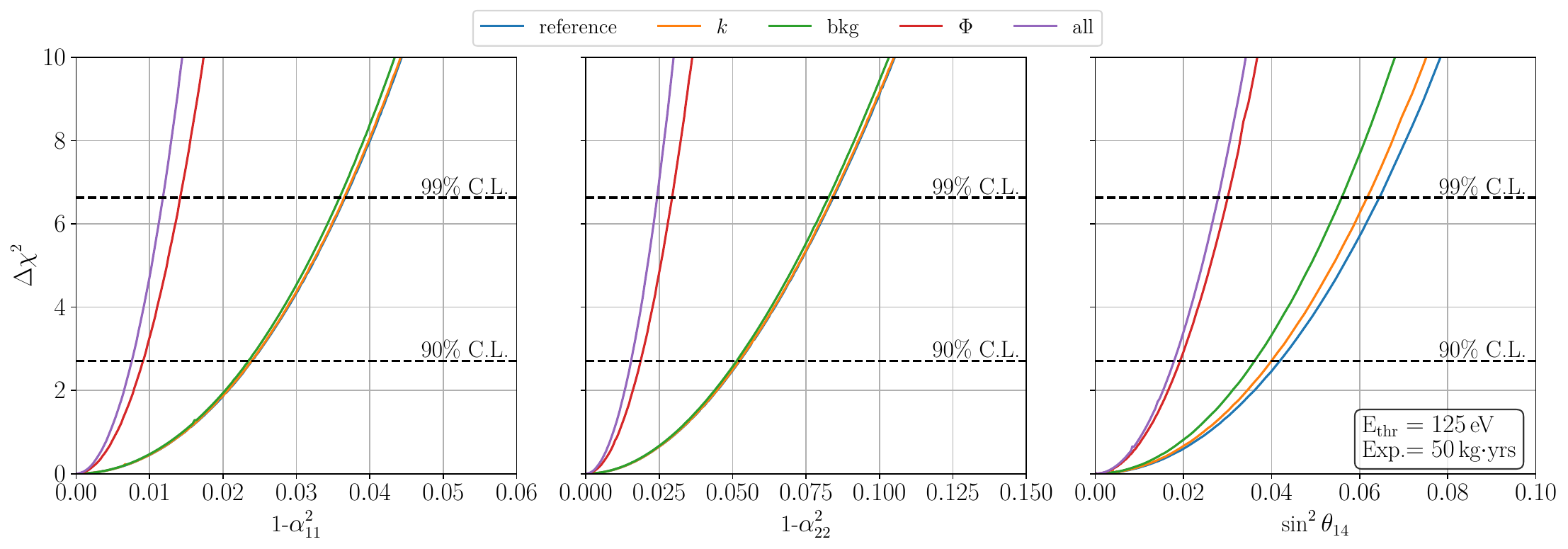}
        \caption{$\Delta \chi^{2}$ profiles of individual model parameters for improved experimental characteristics: quenching uncertainty reduced by a factor 2 (orange), a factor 10 improvement on the background level (green) and flux uncertainty (red). The blue curve shows our reference configuration, while the purple lines indicate the combination of improvements. In the heavy seesaw limit (left and middle), the other alpha parameter is fixed to unity, while for the light sterile case (right) we set $\Delta m^{2}_{14}=1$\,eV$^{2}$ for illustrative purposes. As example a 125\,eV-threshold detector with 50\,kg$\cdot$yr exposure is chosen.}
        \label{fig:parameter_improvement} 
\end{figure}


The additional knowledge from oscillation sites - indicated by gray contours in figure \ref{fig:alpha_parameter_space} - strongly shrinks the parameter space still allowed by our CE$\nu$NS setup.
As evident from the single parameter cases, limits on $\alpha_{22}$ are driven by oscillation experiments, while CE$\nu$NS experiments are contributing valuable knowledge for $\alpha_{11}$, underlying their importance for global investigations.

The optimized configuration (a factor 10 improvement in flux uncertainty and background level as well as a factor 2 improvement on quenching parameter $k$) is able to constrain large parts of the parameter space.
In particular, the transition from 5\,kg$\cdot$yrs to 50\,kg$\cdot$yrs shows a strong increase in sensitivity regardless of the chosen detection threshold.
Note also that these systematics are still not limiting because improved thresholds and exposures lead to better limits on the non-unitarity parameters.

In order to quantify which experimental parameters are the main drivers of the improved sensitivity, we perform sensitivity estimates switching them on one after the other.
Our findings are summarized in figures \ref{fig:parameter_improvement} for single parameters and \ref{fig:alpha_parameter_space_optimization} and \ref{fig:light_sterile_parameter_space_optimization} in the case of two parameters and both mass regimes.
It is apparent that the uncertainty of the reactor antineutrino flux is the limiting factor, leading to a relative improvement of $\sim63\%$ for the constraints on the non-unitarity parameters when reduced by a factor 10.
Reducing the background level by a factor 10 only improves the limits by $\sim2\%$.
Further knowledge on the $k$ parameter of the Lindhard model seems to be of minor importance, i.e.\ $<0.5\%$.
Finally, an overall improvement of constraints on the alpha parameters of $\sim70\%$ can be achieved when all factors are combined. 
Nevertheless, it is clear that better knowledge about reactor-related quantities will become the main limitation for the next-generation of experiment.


\subsection{Light sterile limit}

\begin{figure}[H]
\centering
    \includegraphics[width=0.49\textwidth]{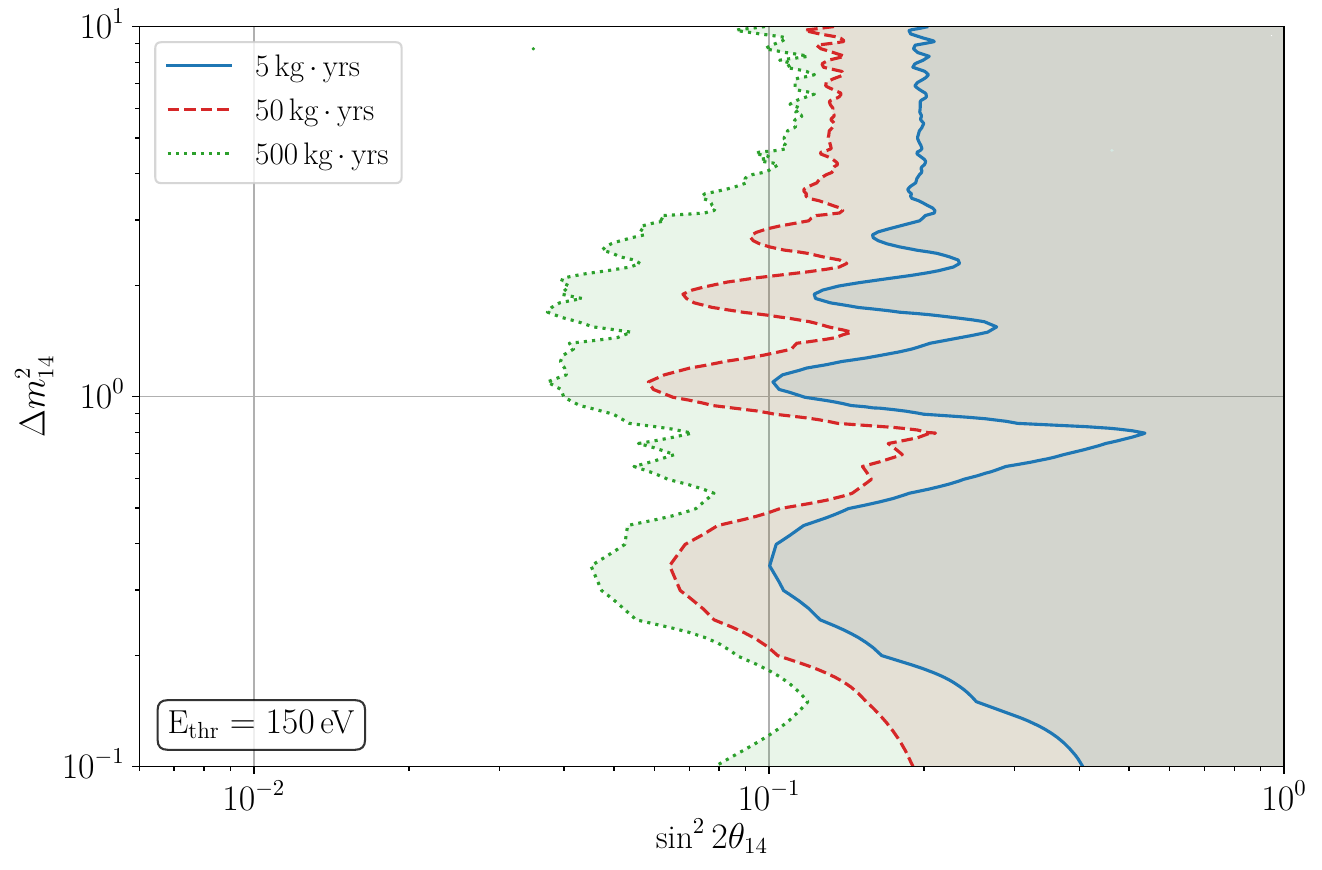}
    \hfill
    \includegraphics[width=0.49\textwidth]{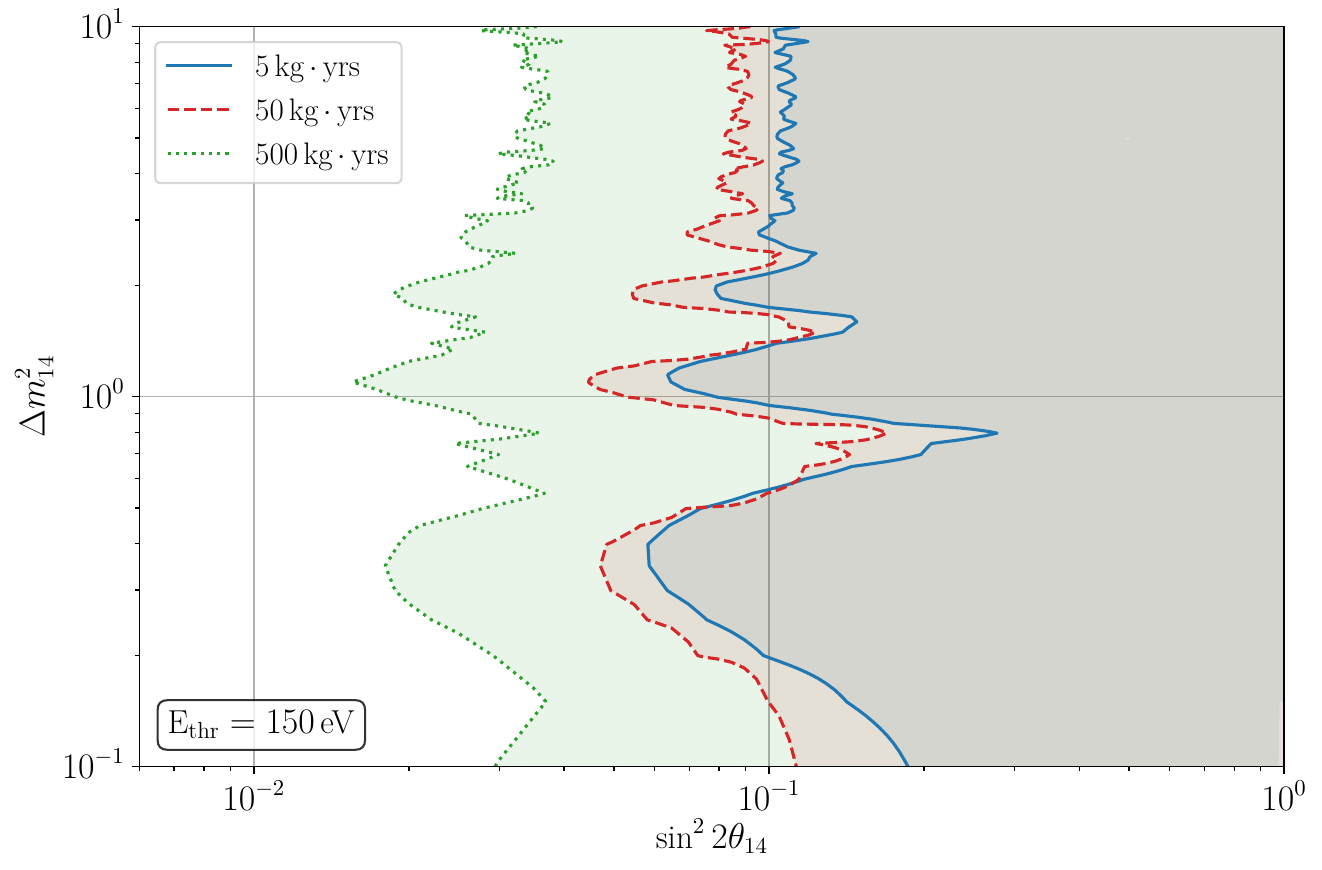}\\
    \vspace{1cm}
    \includegraphics[width=0.49\textwidth]{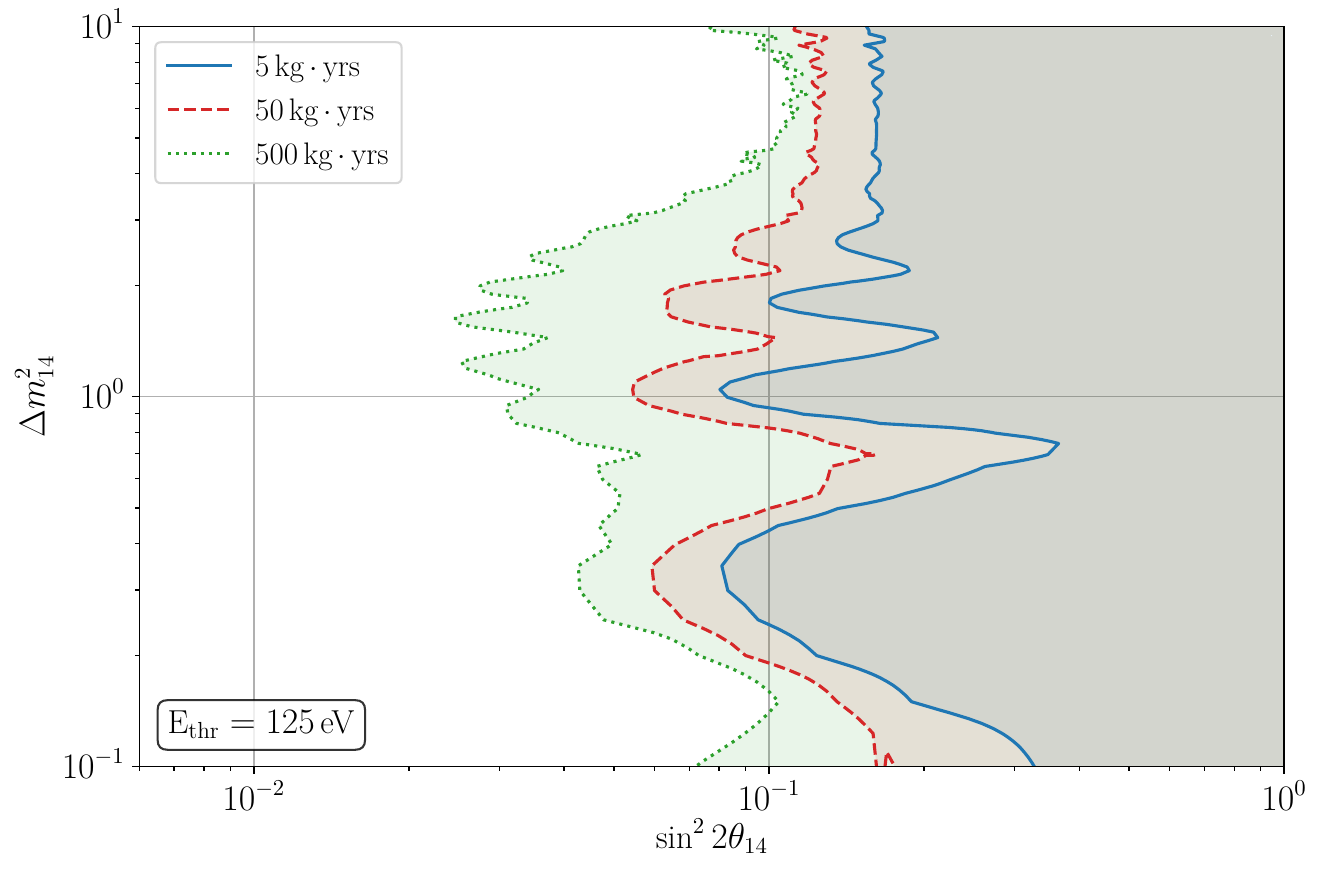}
    \hfill
    \includegraphics[width=0.49\textwidth]{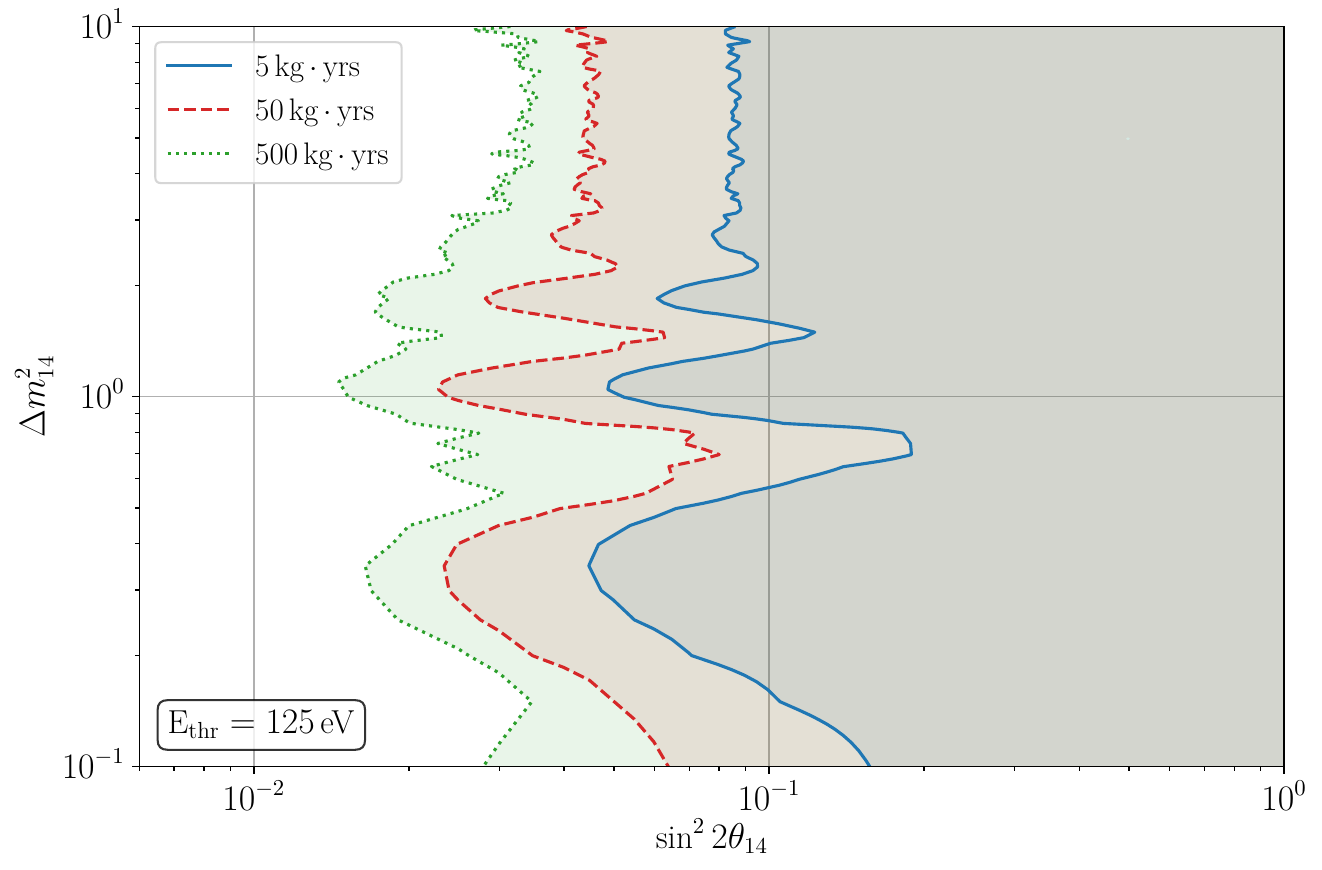}\\
    \vspace{1cm}
    \includegraphics[width=0.49\textwidth]{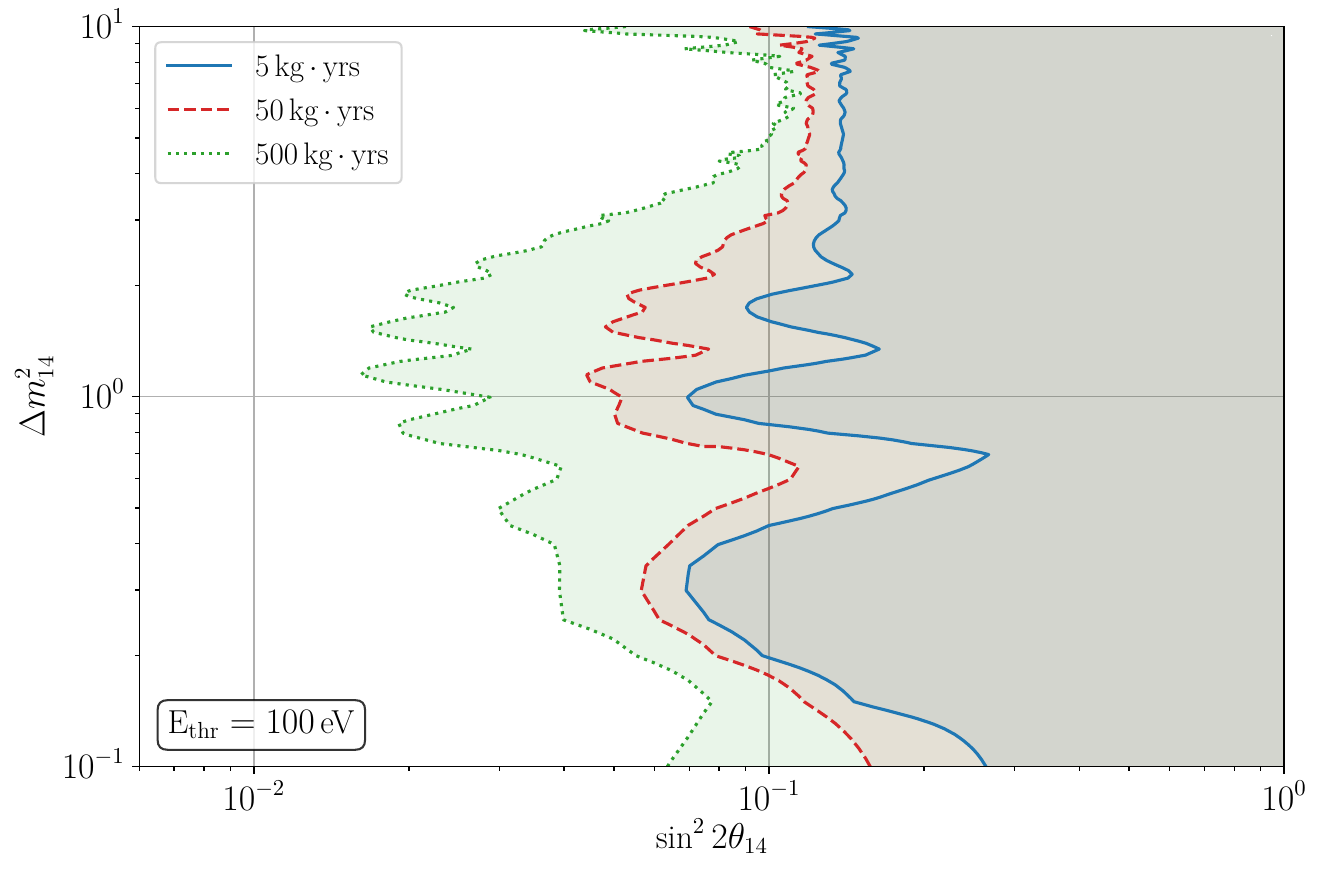}
    \hfill
    \includegraphics[width=0.49\textwidth]{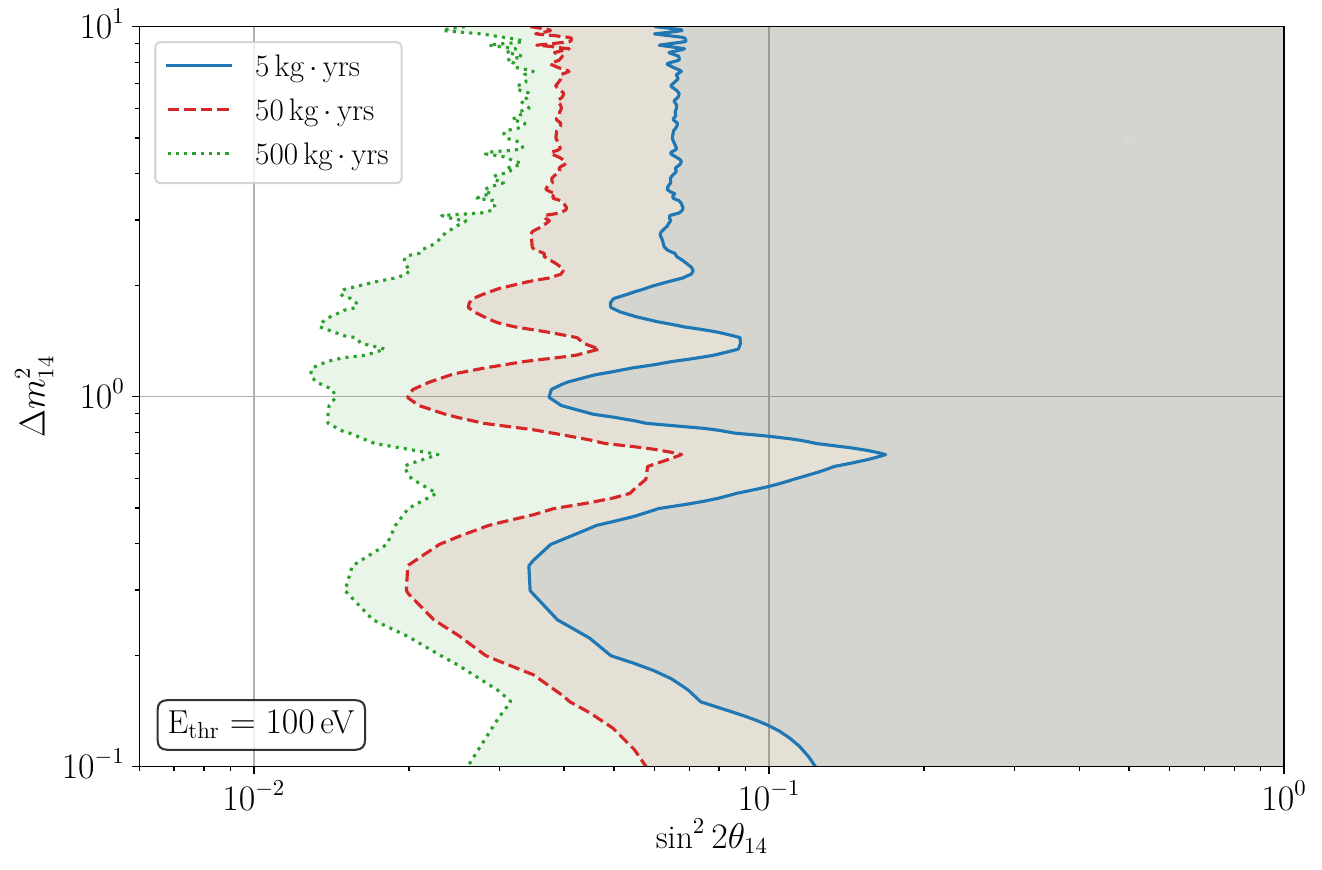}
    \caption{Experimental sensitivity (exclusion potential) of light sterile neutrino searches of our reference setup (left column) with the assumed detection thresholds (150, 100, 50)\,eV and exposures (5, 50, 500)\,kg$\cdot$yrs. In contrast to the previous case, such searches are not limited by the systematic uncertainties assumed. Results for improved experimental characteristics (a factor 10 improvement in flux uncertainty and background level as well as a factor 2 improvement in quenching uncertainty) are shown as well (right columns). Descriptions of all configurations are given in Tab.~\ref{tab:benchmarks}.}
\label{fig:light_sterile_parameter_space} 
\end{figure}

Similar to the previous case, we expect CE$\nu$NS to have the dominating contribution to the achieved sensitivity, cf.\ figure \ref{fig:single_channel_light_sterile} for exemplary single channel sensitivities.
In addition, it is worth noting that there already exist many dedicated experiments that aim to investigate a light sterile neutrino, especially at sites very close to a nuclear reactor core \cite{PROSPECT:2020sxr,STEREO:2019ztb,DANSS:2018fnn,NEOS:2016wee,Serebrov:2021ndx}.
Our sensitivity results are illustrated in figure \ref{fig:light_sterile_parameter_space}. 
We immediately see that in the light sterile limit systematic uncertainties are not limiting since higher exposures and lower detection thresholds still show constraining power.
For example, a detector of 50\,kg$\cdot$yr exposure and a threshold of 125\,eV mostly excludes mixing angles $\sin^{2}2\theta_{14}\gtrsim0.2$ for $\Delta m^{2}_{14} \in [10^{-1}, 10]\,$eV$^{2}$. 
The BEST experiment best-fit \cite{Barinov:2022wfh}, $\Delta m^2 = 3.3^{+\,\infty}_{-2.3}\,\text{eV}^2$ and $\sin^2 2\theta = 0.42^{+0.15}_{-0.17}$, is fully excluded in our projections, a result consistent with the strong tension between the BEST anomaly and other short-baseline data \cite{Giunti:2022btk}.
Moreover, 100\,eV-detectors with 500\,kg$\cdot$yr exposure will start to probe mixing angles $\sin^{2}2\theta_{14}\lesssim2\cdot10^{-2}$ in a setup at 20\,m-distance.
While in the previous case, configuration of almost similar sensitivity could be identified, i.e.\ 500\,kg$\cdot$yrs at 150\,eV vs.\ 50\,kg$\cdot$yrs at 100\,eV, the message here is different:
CE$\nu$NS searches for a light sterile neutrino clearly benefit from a lower detection threshold. 

Improved experimental specifications, i.e.\ antineutrino flux, background level and quenching, also boost the experimental sensitivity in this context. 
In particular, a setup with a threshold below 125~eV and an exposure larger than 50\,kg$\cdot$yrs will clearly exclude mixing angles above 0.1, and larger parts above 0.05.

To quantify the effect of experimental uncertainties also for this case, we fixed $\Delta m^{2}=1$\,eV$^{2}$ and varied them individually.\footnote{Here we have selected a generic point close to the second main oscillation peak. Of course, the results vary depending on the chosen value of $\Delta m^{2}$ and the chosen experimental characteristics.}
The $\Delta \chi^{2}$ profiles in terms of the mixing angle $\sin^{2}\theta_{14}$ are given in figure \ref{fig:parameter_improvement} (right plot).
Also here, the flux uncertainty is the driving factor of the obtained sensitivity with $\sim54\%$ relative improvement. 
The impacts of quenching and background level are stronger, i.e.\ $\sim4\%$ and $\sim14\%$ relative improvements, but are still of secondary importance.
An overall improvement of $\sim 57\%$ can be gained with the combination of these factors, less strong than for the seesaw limit.
The effects on the two-dimensional parameter space are illustrated in figure \ref{fig:light_sterile_parameter_space_optimization}.

However, the parameter space probed is mostly excluded by existing short-baseline experiments. 
The advantage of CE$\nu$NS setups in this context lies in their compactness.
An experiment of several medium-size CE$\nu$NS detectors at different distances would allow to reduce reactor-related uncertainties.
In addition, CE$\nu$NS is sensitive to all neutrino flavors, providing complementary information to charged current electron-flavor–based searches.


\begin{table}[t]
\begin{tabular}{|l|c|c|}
\hline
 & $M$ limit from $\alpha_{11}$ & $M$ limit from $\alpha_{22}$ \\
\hline
oscillations & $700$\,GeV & $1740$\,GeV \\
oscillations + now & $1300$\,GeV & $1940$\,GeV \\
oscillations + soon & $1330$\,GeV & $1940$\,GeV \\
oscillations + future & $2460$\,GeV & $2250$\,GeV \\
\hline
\end{tabular}
\caption{Limits (at 90\% C.L.) on the heavy mediator mass scale $M$ inferred from the 
non-unitarity parameters $\alpha_{11}$ and $\alpha_{22}$, assuming a low-scale  seesaw with $\mathcal{O}(1)$ coefficients. The row ``oscillations'' uses current bounds from Ref.~\cite{Forero:2021azc}. 
The rows ``oscillations + current/realistic/optimistic'' additionally include projected CE$\nu$NS sensitivities for CONUS-like detector thresholds and exposures of 150\,eV with 5\,kg\,yr (now), 125\,eV with 50\,kg\,yr (soon), and 100\,eV with 500\,kg\,yr (future). In the optimistic case we assume the optimized experimental configuration, see the main text for more details.}
\label{tab:Mlimits}
\end{table}

\section{Conclusions}\label{sec:conclusion}

\noindent After the first detection of CE$\nu$NS, more and more CE$\nu$NS experiments start data collection and we can expect further interesting physics results in the future, not only from experiments but also from subsequent phenomenological studies.
The existing data and analyses already revealed the large potential of CE$\nu$NS that could be further exploited when the next generation of experiments transition to precision measurements.
Motivated by the recent CE$\nu$NS observation of the CONUS+ experiment, the present work aimed at identifying the sensitivity of a future upgrade based on demonstrated technology, i.e.\ a next-generation Germanium-based experiment close to a nuclear power reactor site.
With lepton non-unitarity as example, the expected experimental reach has been determined for well-selected benchmark points, while taking into account major experimental uncertainties such as reactor antineutrino flux and signal quenching.
Further, the impact of these uncertainties and the underlying background level in future experimental endeavors has been assessed to identify key drivers for scientific progress in this context.

In the so-called seesaw limit, where new degrees of freedom are heavy, a future CONUS-like experiments will contribute valuable information on the non-unitarity parameters, cf.\ table \ref{tab:single_alpha_limits} and figures \ref{fig:alpha_profiles} and \ref{fig:alpha_parameter_space}.
A CE$\nu$NS setup with characteristics soon to be achieved (50\,kg$\cdot$yr, 125\,eV-threshold) will be able to probe non-unitarity parameters related to energy scales of 1100GeV (for $\alpha_{11}$) and 760GeV (for $\alpha_{22}$).
With reduced uncertainties on the reactor antineutrino flux (factor 10), quenching (factor 2) and a reduced background level (factor 10) scales up to 2500\,GeV (for $\alpha_{11}$) and 1700\,GeV (for $\alpha_{22}$) could be tested in a future setup. In general, when combined with knowledge from oscillation experiments even higher scales can be probed, see table \ref{tab:Mlimits}. We emphasize that the mass scales inferred from Eq.~\ref{eq:limit_to_mass} are not strict experimental limits: they rely on the assumptions specified in the main text, including the low-scale seesaw interpretation and $\mathcal{O}(1)$ coefficients.

For the case of light new particle (light sterile limit), bounds from CE$\nu$NS experiment will be improved significantly, cf.\ figure \ref{fig:light_sterile_parameter_space}.
Especially, mixing angles above $\sim0.1$ could be fully excluded when systematic uncertainties and background can be further lowered.
Obtained results are not competitive to existing bounds, but are exceptional in the sense that they are flavor-independent.
Further, a more refined investigation with several CE$\nu$NS detectors may be envisioned.
For future design, our work contributes interesting knowledge since experimental configurations with (roughly) the same sensitivity have been identified.
In particular, the question whether to build an experiment with 500\,kg$\cdot$yr exposure and a 150eV-threshold or 50\,kg$\cdot$yr with a 100\,eV-threshold may be answered by detector developments in the next years.
Furthermore, it has become clear that systematic uncertainties will be the limiting factor in future precision experiments underlining the importance of carefully assessing an experiment's uncertainties and improved theory predictions.
In the context of this work, the reactor antineutrino flux is identified as one of the key drivers for experimental sensitivity, cf.\ figure \ref{fig:parameter_improvement}.
Of course, more refined studies from our experimental colleagues are needed to fully consider all potential uncertainties underlying a specific experimental setup.

Nevertheless, our work clearly underlines the strong potential of future CE$\nu$NS experiments for future tests of the lepton sector and searches of physics beyond the standard model.

\section*{Acknowledgments}

Work supported by the Spanish grants PID2023-147306NB-I00 and CEX2023-001292-S (MICIU/AEI/ 10.13039/501100011033), as well as CIPROM/2021/054 (Generalitat Valenciana). TR and SCCh acknowledge support by the MPIK Heidelberg, where this work started. The authors would like to thank Edgar Sánchez García for discussions on realizing the experimental configurations investigated in this work.
\appendix


\bibliography{biby}
\bibliographystyle{utphys}


\appendix
\section{Detailed $\Delta\chi^{2}$ profiles of the seesaw limit}

\begin{figure}[H]
\centering
    \includegraphics[height=0.29\textheight]{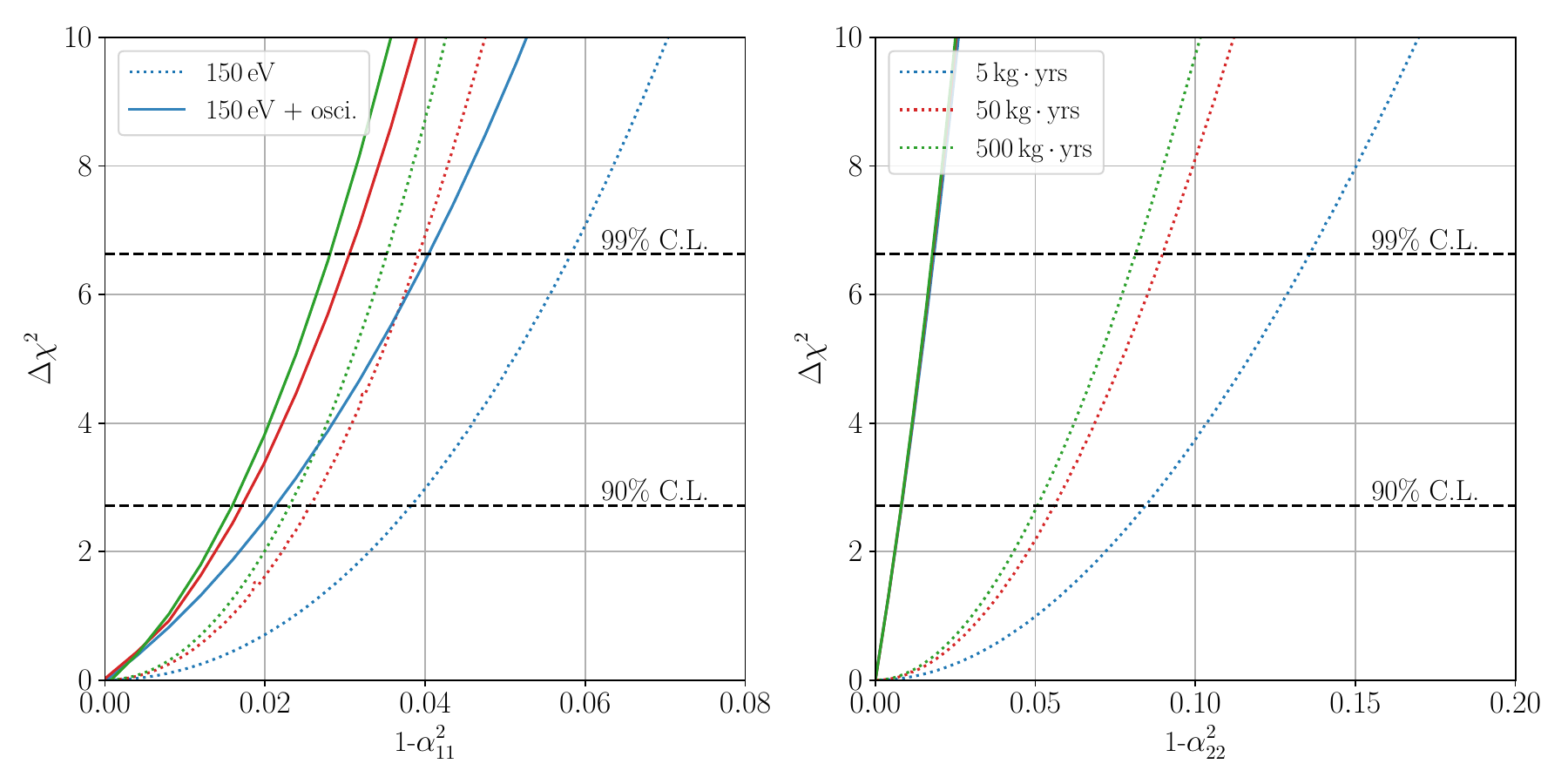}\\ 
    \includegraphics[height=0.29\textheight]{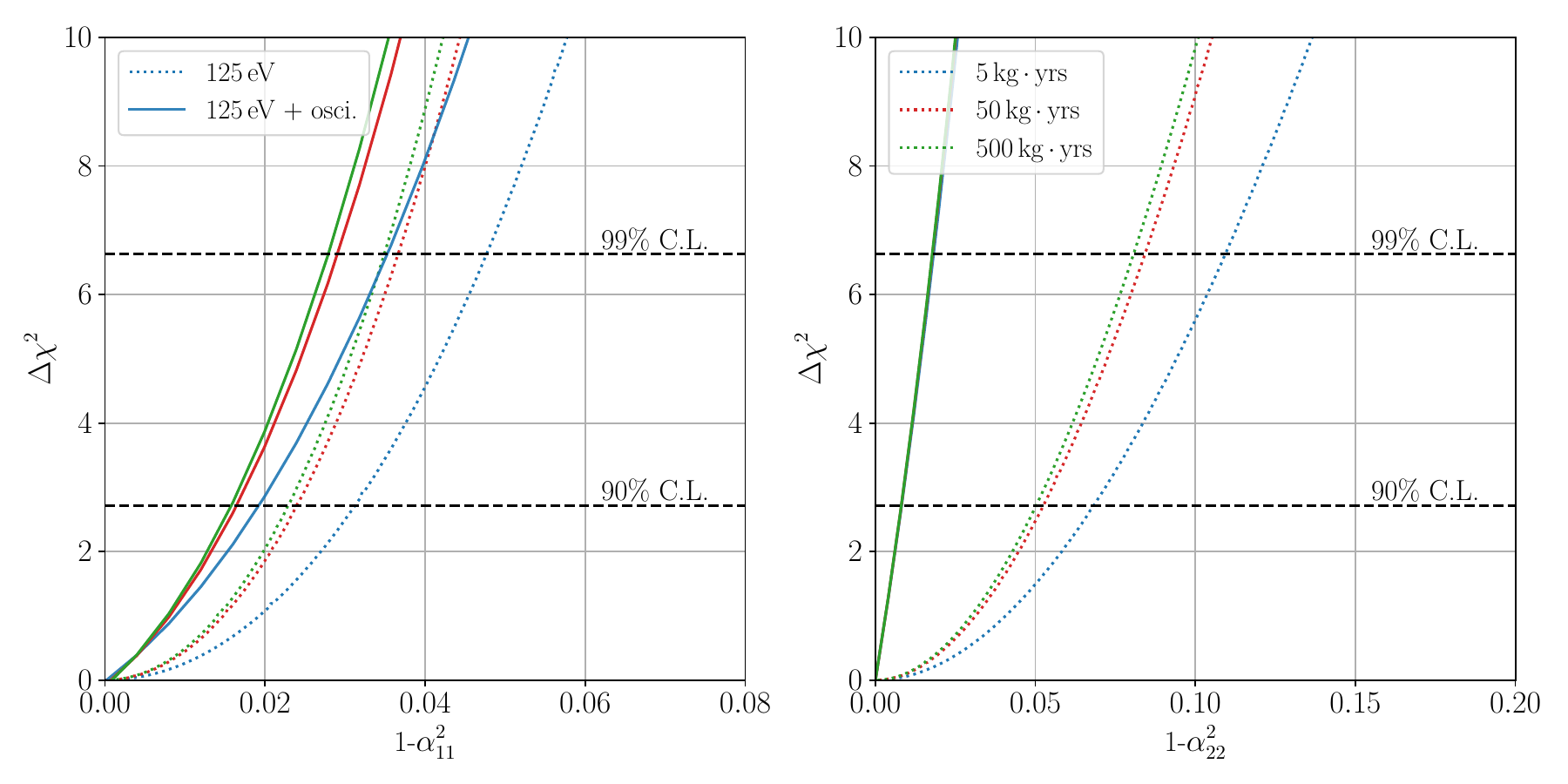}\\
    \includegraphics[height=0.29\textheight]{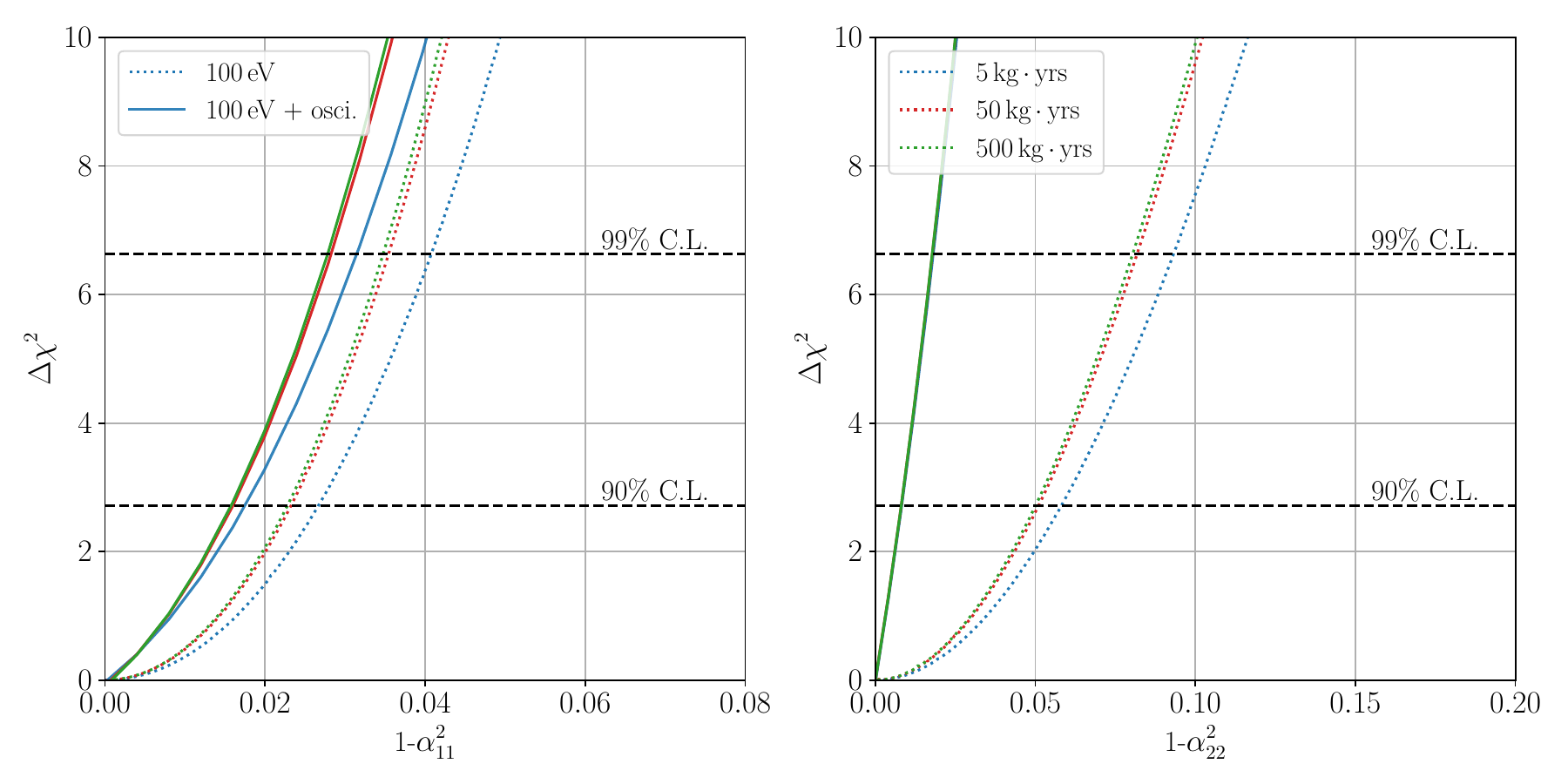} 
    \caption{Detailed $\Delta \chi^{2}$ profiles for the individual alpha parameters and the three threshold values (150\,eV, 125\,eV, 100\,eV from top to bottom) under consideration. The experimental reach of the individual experimental configuration (dashed lines) is shown in comparison to the sensitivity when information from oscillation experiments is added as external knowledge to our analysis (solid). Benchmark configurations are described in Tab.~\ref{tab:benchmarks}.}
\label{fig:alpha_profiles_detailed} 
\end{figure}

\section{Impact of improved experimental parameters - full parameter space}


\begin{figure}[H]
    \centering
    \includegraphics[width=0.99\textwidth]{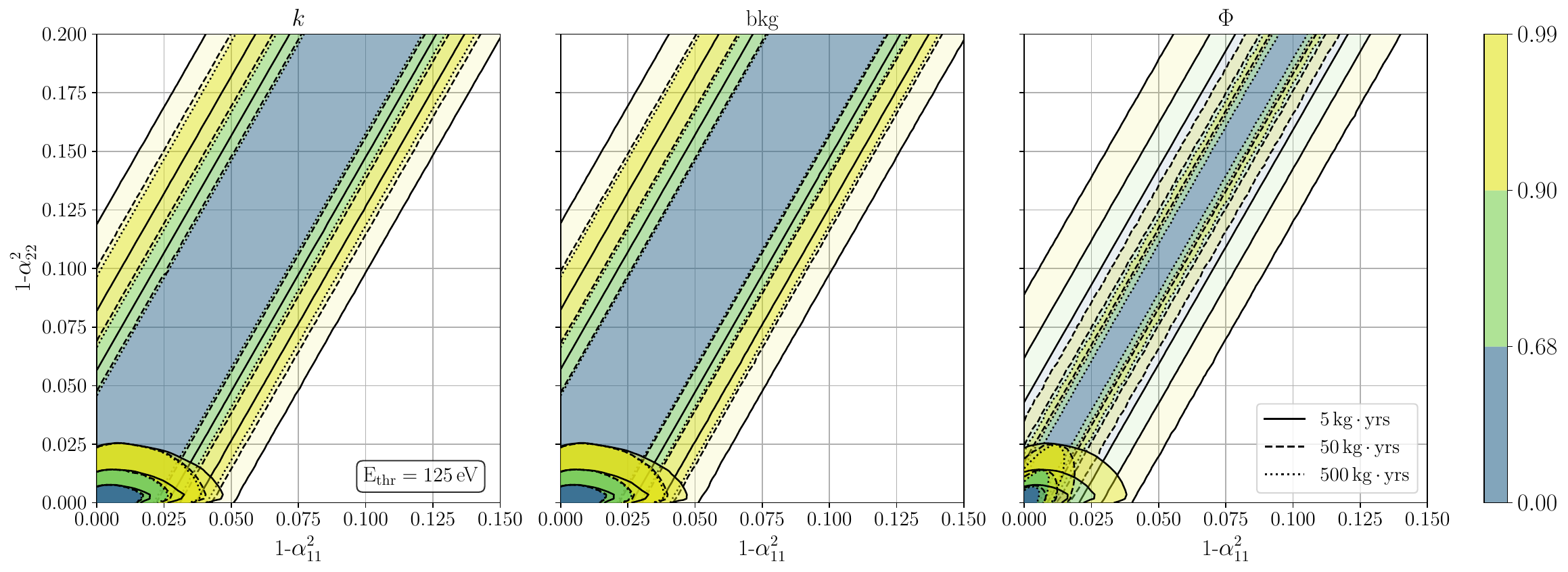} 
    \caption{Limits on the non-unitarity parameters under optimization of experimental characteristics: quenching uncertainty reduced by a factor 2 (left), background reduction by a factor of 10 (middle) and antineutrino flux uncertainty reduced by a factor 10 (right). Here, a 125eV-threshold detector with 50\,kg$\cdot$yr exposure is chosen as example.}
\label{fig:alpha_parameter_space_optimization} 
\end{figure}

\begin{figure}[H]
\centering
\includegraphics[width=0.99\textwidth]{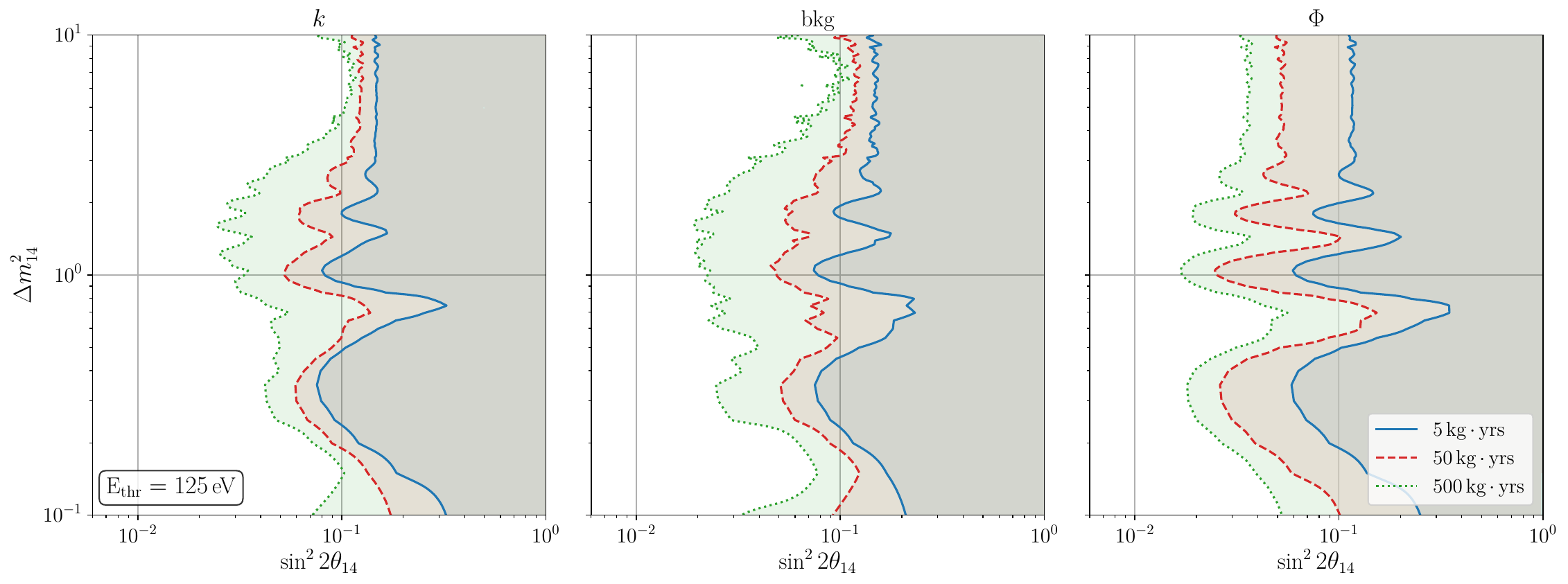}
\caption{Limits on mass and mixing parameters under optimization of experimental characteristics: quenching uncertainty reduced by a factor 2 (left), background reduction by a factor of 10 (middle) and antineutrino flux uncertainty reduced by a factor 10 (right). Here, a 125\,eV-threshold detector with 50\,kg$\cdot$yr exposure is chosen as example.}
\label{fig:light_sterile_parameter_space_optimization} 
\end{figure}



\section{Single channel comparison}

\subsection{Seesaw limit}

\begin{figure}[H]
    \includegraphics[width=0.5\textwidth]{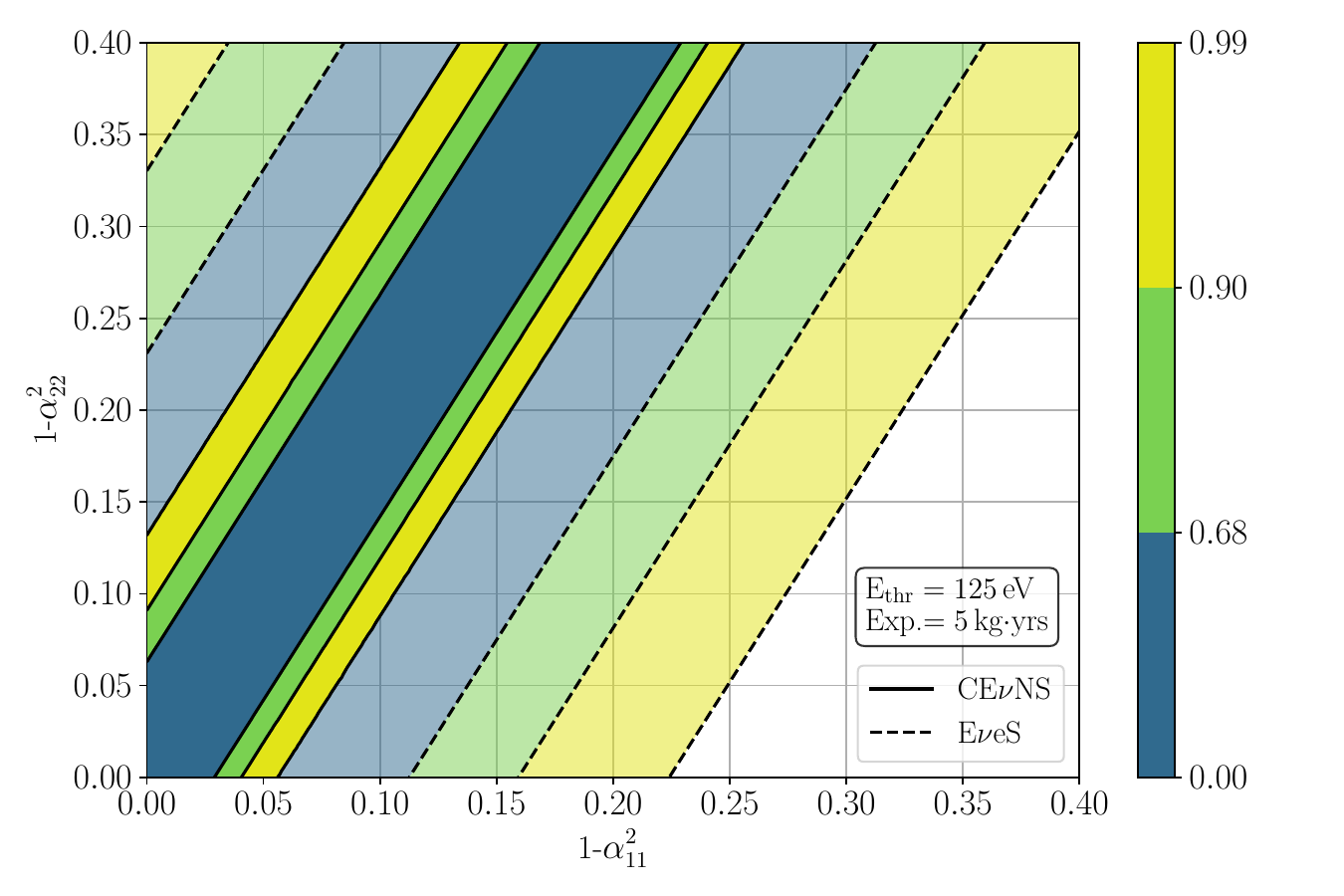}\hfill
    \includegraphics[width=0.5\textwidth]{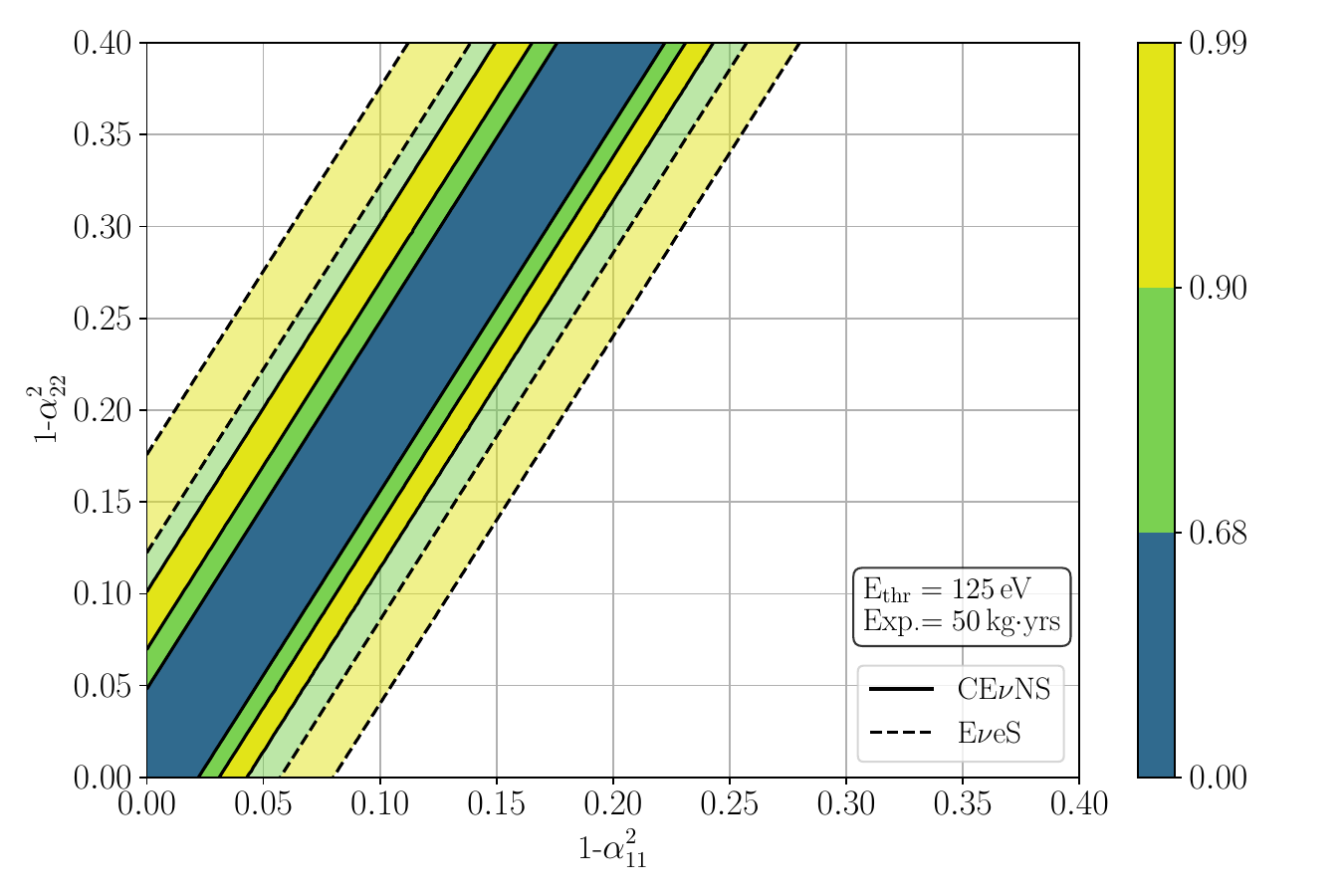}
    \caption{Sensitivity of CE$\nu$NS and E$\nu$eS for the assumed experimental configuration. Limits (at 90\% C.L.) on the alpha parameters are given for CE$\nu$NS below 1\,keV and E$\nu$eS above 1\,keV up to 100\,keV. It is evident that the obtained results are dominated by CE$\nu$NS.}
\label{fig:single_channel_alpha_parameter} 
\end{figure}

\subsection{Light sterile limit}

\begin{figure}[H]
    \includegraphics[width=0.5\textwidth]{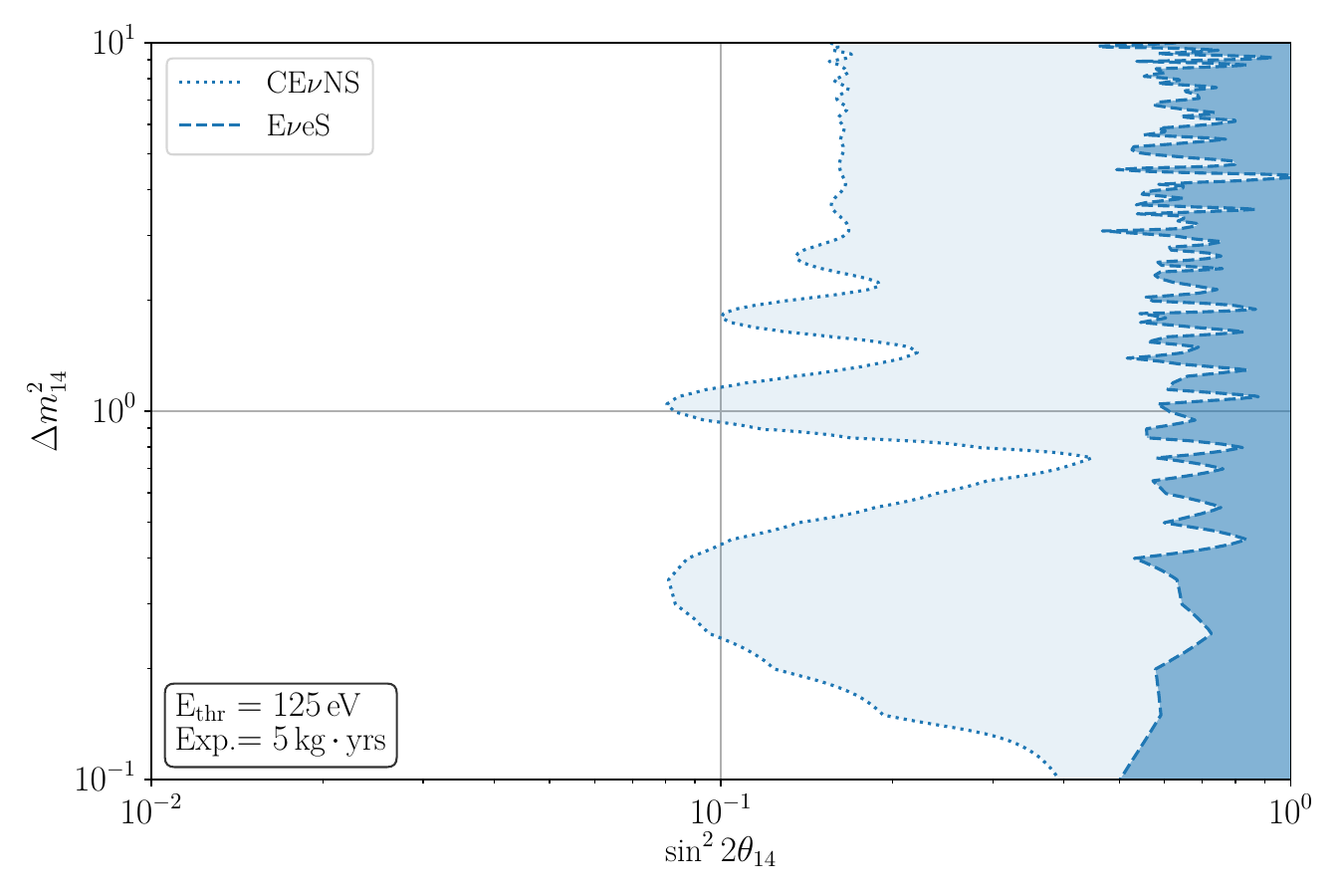}\hfill
    \includegraphics[width=0.5\textwidth]{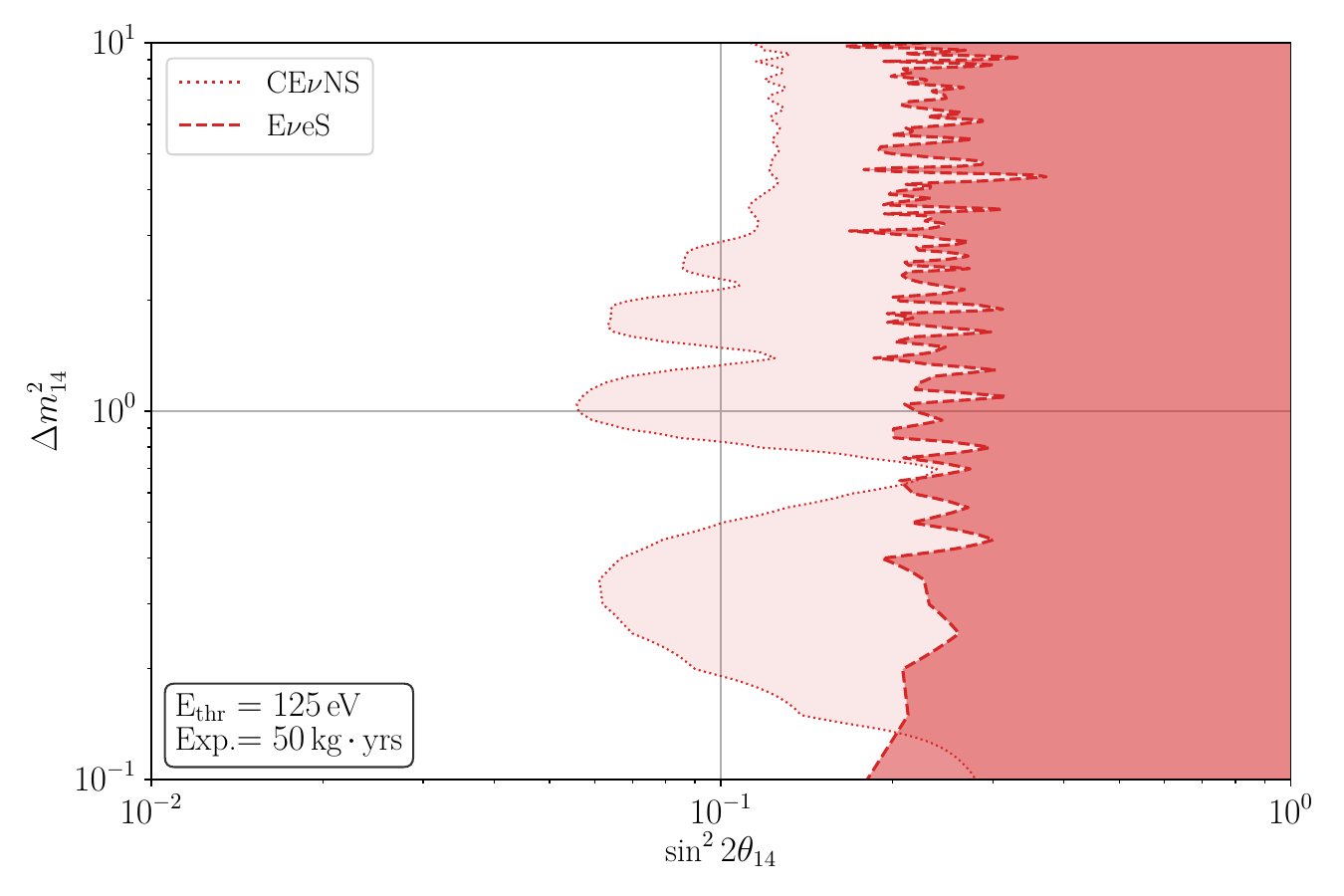}
    \caption{Exclusion potential of CE$\nu$NS and E$\nu$eS for the assumed experimental configuration. Limits (at 90\% C.L.) on the mixing angle and the mass-squared difference are given for CE$\nu$NS below 1\,keV and E$\nu$eS above 1\,keV up to 100\,keV. Also here, CE$\nu$NS is the determining factor for the expected sensitivity.}
\label{fig:single_channel_light_sterile} 
\end{figure}

\end{document}